\documentclass[12pt]{article}
\usepackage{amsfonts,amssymb,mathrsfs}
\usepackage{hyperref}
\newcommand{\beq}{\begin{equation}}
\newcommand{\eeq}{\end{equation}}
\begin{document}
\begin{center}
{\Large\bf $D=3$ $\mathcal N=6$ superconformal symmetry\\ of $AdS_4\times\mathbb{CP}^3$ superstring}\\[0.5cm]
{\large D.V.~Uvarov\footnote{E-mail: d\_uvarov@\,hotmail.com, uvarov@\,kipt.kharkov.ua}}\\[0.2cm]
{\it NSC Kharkov Institute of Physics and Technology,}\\ {\it 61108 Kharkov, Ukraine}\\[0.5cm]
\end{center}
\begin{abstract}
Invariance of the $AdS_4\times\mathbb{CP}^3$ superstring under $D=3$ $\mathcal N=6$ superconformal symmetry is discussed in the sector described by the $OSp(4|6)/(SO(1,3)\times U(3))$ supercoset sigma-model action presented in the conformal basis for the $osp(4|6)/(so(1,3)\times u(3))$ Cartan forms. Transformation rules under $D=3$ $\mathcal N=6$ superconformal symmetry for the $(10|24)-$dimensional 'reduced' $AdS_4\times\mathbb{CP}^3$ superspace coordinates are obtained and used to derive corresponding world-sheet currents.
\end{abstract}
\setcounter{equation}{0}
\def\theequation{\thesection.\arabic{equation}}

\section{Introduction}
The first explicit example of the gauge/string duality \cite{Maldacena97}, \cite{GKP98}, \cite{Witten98} allowed to  probe analytically previously unaccessible nonperturbative regime of $\mathcal N=4$ super-Yang-Mills theory via the $IIB$ superstring on $AdS_5\times S^5$ background. It is in a sense the simplest instance of the $AdS/CFT$ correspondence due to the maximal supersymmetry described by $PSU(2,2|4)$ supergroup both of the $AdS_5\times S^5$ superbackground and $D=4$ supersymmetric conformal field theory (SCFT) on the boundary of $AdS_5$ space.

Another highly supersymmetric explicit example of the $AdS/CFT$ correspondence that was put forward not long ago by Aharony, Bergman, Jafferis and Maldacena (ABJM) \cite{ABJM} provides dual description of the SCFT in the space-time of one lower dimension in terms of $M-$theory on $AdS_4\times(S^7/\mathbb{Z}_k)$ background. In spite of the fact that lower dimensional theories basically have simpler dynamics compared to $4d$ ones the ABJM correspondence appears to be harder to verify since on both sides of the duality the isometry supergroup $OSp(4|6)$ isomorphic to $D=3$ $\mathcal N=6$ superconformal symmetry is lower than the maximally allowed one.

Difficulties manifest itself already at the level of constructing the classical action for the $IIA$ superstring on $AdS_4\times\mathbb{CP}^3$ superbackground that provides dual description of the 't Hooft limit of $D=3$ SCFT proposed by ABJM \cite{ABJM}. Group-theoretic supercoset approach \cite{MT98}, \cite{KalloshRajaraman98}, \cite{BBHZ}, \cite{RoibanSiegel} originally elaborated to describe the $IIB$ superstring on $AdS_5\times S^5$ background when applied to the $AdS_4\times\mathbb{CP}^3$ superstring gives the partial answer \cite{AF08}, \cite{Stefanski}\footnote{Alternative way to construct the
$AdS_4\times\mathbb{CP}^3$ superstring using the pure spinor approach was followed in \cite{PS}.} because only the subspace of $AdS_4\times\mathbb{CP}^3$ superspace can be realized as the supercoset manifold $OSp(4|6)/(SO(1,3)\times U(3))$. The $OSp(4|6)/(SO(1,3)\times U(3))$ sigma-model action \cite{AF08}, \cite{Stefanski} corresponds to fixing half of the gauge freedom related to $\kappa-$symmetry of the complete action \cite{GSWnew} that can be obtained via the double dimensional reduction \cite{DHIS} of the $D=11$ supermembrane action on the maximally supersymmetric $AdS_4\times S^7$ background \cite{Plefka98} due to the Hopf fibration realization of the 7-sphere $S^7=\mathbb{CP}^3\times S^1$ \cite{Pope}, \cite{STV85}. Although $OSp(4|6)/(SO(1,3)\times U(3))$ supercoset sigma-model fails to describe all possible $AdS_4\times\mathbb{CP}^3$ superstring configurations \cite{AF08}, \cite{GSWnew} it has clear group-theoretical structure and is classically integrable allowing one to utilize for its investigation many of the results obtained for the "elder brother" example of $AdS_5/CFT_4$ correspondence relying on the intregrable structures\footnote{The issue of seeking possible integrable structures for the $AdS_4\times\mathbb{CP}^3$ superstring beyond the $OSp(4|6)/(SO(1,3)\times U(3))$ sigma-model has been recently addressed in Ref.\cite{SW1009}.} exhibited there \cite{Zarembo}, \cite{BPR} (for the collection of recent reviews see, e.g., \cite{JPA}).

In Ref. \cite{U08} we have found explicit form of the $OSp(4|6)/(SO(1,3)\times U(3))$ supercoset sigma-model action in the conformal basis for $osp(4|6)$ Cartan forms considering the supercoset representative parametrized by Poincare coordinates for the $AdS_4$ space with 24 fermionic coordinates split into two sets of 12 related to Poincare and conformal supersymmetries from the $AdS$ boundary superspace perspective\footnote{Previously such conformal-type parametrizations were used to examine the string/brane models involved into the higher-dimensional examples of $AdS/CFT$ correspondence \cite{9807115}, \cite{Kallosh2}, \cite{PST}, \cite{MT2000}.}. Such a choice of the supercoset representative allows to formulate the stringy side of the duality in terms of the variables that contain those parametrizing $D=3$ $\mathcal N=6$ boundary superspace, where the ABJM theory could be formulated off-shell \cite{Benna}, \cite{Cederwall}, \cite{Buchbinder} aiming at getting new insights into the relation between both theories.

The goal of this paper is to establish transformation properties under $D=3$ $\mathcal N=6$ superconformal symmetry of the $OSp(4|6)/(SO(1,3)\times U(3))$ supercoset coordinates introduced in Ref. \cite{U08}, as well as to find Noether currents associated with the $D=3$ $\mathcal N=6$ superconformal invariance of the $OSp(4|6)/(SO(1,3)\times U(3))$  superstring action. Similar problem of deriving $D=4$ $\mathcal N=4$ superconformal transformations for the $AdS_5\times S^5$ superspace coordinates relevant to the $AdS_5/CFT_4$ correspondence was addressed in \cite{Claus}, \cite{Robins}. We start with reviewing the $OSp(4|6)/(SO(1,3)\times U(3))$  sigma-model in the conformal basis for Cartan forms, then examine the action of left $D=3$ $\mathcal N=6$ superconformal transformations on the $OSp(4|6)/(SO(1,3)\times U(3))$ supercoset element and proceed to derivation of the $osp(4|6)$ Cartan forms transformation rules and Noether current densities for each of the individual transformations from $D=3$ $\mathcal N=6$ superconformal symmetry.

\section{$OSp(4|6)/(SO(1,3)\times U(3))$ superstring in conformal basis}

The sigma-model action on the $(10|24)-$dimensional $OSp(4|6)/(SO(1,3)\times U(3))$ superspace was found in \cite{AF08}, \cite{Stefanski} following the general prescription \cite{MT98}, \cite{KalloshRajaraman98}, \cite{BBHZ}, \cite{RoibanSiegel} for constructing sigma-model-type actions on supercoset spaces that admit 4-element outer automorphism $\mathbb{Z}_4$ of the underlying isometry superalgebra. It relies on identifying Cartan forms associated with 10 bosonic and 24 fermionic generators of the $osp(4|6)/(so(1,3)\times u(3))$ supercoset as the $(10|24)-$dimensional supervielbein components. Resulting action is invariant under the global $OSp(4|6)$ supersymmetry, as well as gauge $SO(1,3)\times U(3)$ and $\kappa-$symmetries, describes the requisite number of physical degrees of freedom, has correct bosonic limit and is classically integrable.

In Ref.\cite{U08} we have considered the $OSp(4|6)/(SO(1,3)\times U(3))$ supercoset element
\begin{equation}\label{scosetelement}
\mathscr G=e^{x^mP_m+\theta^\mu_aQ^a_\mu+\bar\theta^{\mu a}\bar Q_{\mu a}}e^{\eta_{\mu a}S^{\mu a}+\bar\eta^a_\mu\bar S^{\mu}_a}e^{z^aV_a{}^4+\bar z_aV_4{}^a}e^{\varphi D}
\end{equation}
parametrized by $D=3$ $\mathcal N=6$ super-Poincare coordinates $(x^m, \theta^\mu_a, \bar\theta^{\mu a})$, $AdS_4$ radial direction coordinate $\varphi$ related to the boundary-space dilatations, 3 complex coordinates $(z^a, \bar z_a)$ of the $\mathbb{CP}^3$ manifold, and 12 fermionic coordinates $(\eta_{\mu a}$, $\bar\eta^a_\mu)$ corresponding to $D=3$ $\mathcal N=6$ conformal supersymmetry. Associated current 1-form in the conformal basis reads
\beq
\label{defcartan}
\begin{array}{rl}
\mathscr C(d)=\mathscr G^{-1}d\mathscr G=&
\hat\omega^m(d)P_m+\hat
c^m(d)K_m
+\Delta(d)D+G^{mn}(d)M_{mn}\\[0.2cm]
+&\Omega_a{}^4(d)V_4{}^a
+\Omega_4{}^a(d)V_a{}^4+\Omega_a{}^b(d)V_b{}^a+\Omega_4{}^4(d)V_4{}^4\\[0.2cm]
+&\hat\omega^\mu_a(d)Q^a_\mu+\hat{\bar\omega}{}^{\mu a}(d)\bar Q_{\mu a}+\hat\chi_{\mu
a}(d)S^{\mu a}+\hat{\bar\chi}{}^a_\mu(d)\bar S^\mu_a
\end{array}
\eeq
or manifesting the $\mathbb{Z}_4-$grading
\beq
\mathscr C(d)=\mathscr C_0(d)+\mathscr C_2(d)+\mathscr C_1(d)+\mathscr C_3(d),
\eeq
where
\beq
\begin{array}{rl}
\mathscr C_0(d)=&\frac12(\hat\omega^m(d)-\hat c^m(d))(P_m-K_m)+G^{mn}(d)M_{mn}+\Omega_a{}^b(d)V_b{}^a+\Omega_4{}^4(d)V_4{}^4,\\[0.2cm]
\mathscr C_2(d)=&\frac12(\hat\omega^m(d)+\hat c^m(d))(P_m+K_m)+\Delta(d)D+\Omega_a{}^4(d)V_4{}^a
+\Omega_4{}^a(d)V_a{}^4,\\[0.2cm]
\mathscr C_1(d)=&\frac12(\hat\omega^\mu_a(d)+i\hat\chi^{\mu}_{a}(d))(Q^a_\mu+iS_{\mu}^{a})+\frac12(\hat{\bar\omega}{}^{\mu a}(d)-i\hat{\bar\chi}{}^{\mu a}(d))(\bar Q_{\mu a}-i\bar S_{\mu a}),\\[0.2cm]
\mathscr C_3(d)=&\frac12(\hat\omega^\mu_a(d)-i\hat\chi^{\mu}_{a}(d))(Q^a_\mu-iS_{\mu}^{a})+\frac12(\hat{\bar\omega}{}^{\mu a}(d)+i\hat{\bar\chi}{}^{\mu a}(d))(\bar Q_{\mu a}+i\bar S_{\mu a}).
\end{array}
\eeq

Then the $\mathbb{Z}_4-$invariant $OSp(4|6)/(SO(1,3)\times U(3))$ superstring action in the conformal basis for Cartan forms (\ref{defcartan}) acquires the form
\beq \label{action}
S=-\frac12\int
d^2\xi\sqrt{-g}g^{ij}\left[\frac14(\hat\omega^m_i+\hat
c^m_i)(\hat\omega_{jm}+\hat
c_{jm})+\Delta_i\Delta_j+\frac12(\Omega_{ia}{}^4\Omega_{j4}{}^a+\Omega_{ja}{}^4\Omega_{i4}{}^a)\right]+S_{WZ}
\eeq
with the Wess-Zumimo action given by
\beq
\begin{array}{rl}
S_{WZ}=&-\frac14\varepsilon^{ij}\int
d^2\xi\left[(\hat\omega^\mu_{ia}+i\hat\chi^\mu_{ia})\varepsilon_{\mu\nu}(\hat{\bar\omega}{}_j^{\nu a}+i\hat{\bar\chi}{}_j^{\nu a})+(\hat\omega^\mu_{ia}-i\hat\chi^\mu_{ia})\varepsilon_{\mu\nu}(\hat{\bar\omega}{}_j^{\nu a}-i\hat{\bar\chi}{}_j^{\nu a})\right]\\[0.2cm]
=&-\frac{1}{2}\varepsilon^{ij}\int
d^2\xi\left(\hat\omega^\mu_{ia}\varepsilon_{\mu\nu}\hat{\bar\omega}{}^{\nu
a}_{j}+\hat\chi_{i\mu
a}\varepsilon^{\mu\nu}\hat{\bar\chi}{}^a_{j\nu}\right).
\end{array}
\eeq
The first two summands in the kinetic part of the action (\ref{action}) include Cartan forms associated with the generators $P_m$ of the space-time translations on the $D=3$ Minkowski boundary of $AdS_4$ space
\beq
\hat\omega^m(d)=e^{-2\varphi}\omega^m(d),\quad\omega^m(d)=dx^m-id\theta^\mu_a\sigma^m_{\mu\nu}\bar\theta^{\nu
a}+i\theta^\mu_a\sigma^m_{\mu\nu}d\bar\theta^{\nu a},
\eeq
conformal boost generators $K_m$
\beq
\begin{array}{rl}
\hat c^m(d)=&e^{2\varphi}c^m(d),\quad
c^m(d)=-id\eta_{\mu
a}\tilde\sigma^{m\mu\nu}\bar\eta^a_\nu+i\eta_{\mu
a}\tilde\sigma^{m\mu\nu}d\bar\eta^a_\nu\\[0.2cm]
+&2(\bar\eta\eta)\left[\eta_{\mu a}\tilde\sigma^{m\mu\nu}(d\bar\theta^a_\nu+\frac14\bar\zeta^a_\nu(d))-(d\theta_{\mu a}+\frac14\zeta_{\mu
a}(d))\tilde\sigma^{m\mu\nu}\bar\eta^a_\nu\right],\quad\bar\eta\eta\equiv\bar\eta^b_\rho\eta^\rho_b,
\end{array}
\eeq
where
\begin{equation}
\zeta^\mu_a(d)=-\tilde\sigma^{m\mu\nu}\omega_m(d)\eta_{\nu
a}=-\tilde\omega^{\mu\nu}(d)\eta_{\nu
a},\quad\bar\zeta^{\mu
a}(d)=-\tilde\sigma^{m\mu\nu}\omega_m(d)\bar\eta_{\nu}^{a}=-\tilde\omega^{\mu\nu}(d)\bar\eta_{\nu}^{a},
\end{equation}
and dilatations
\beq
\Delta(d)=d\varphi+i(d\theta^\mu_a\bar\eta^a_\mu+d\bar\theta^{\mu
a}\eta_{\mu a})
\eeq
with the corresponding generator $D$. Note that the generators $(D, P_m+K_m)$ can be identified as the $so(2,3)/so(1,3)$ coset generators\footnote{(Anti)commutation relations of the $D=3$ $\mathcal N=6$ superconformal algebra can be found in Ref.\cite{U08}.} and corresponding Cartan forms represent the $AdS$ part of the $(10|24)-$supervielbein bosonic components.

Supervielbein components in the directions tangent to the $\mathbb{CP}^3$ manifold are identified with the $su(4)/u(3)$ Cartan forms $(\Omega_a{}^4(d), \Omega_4{}^a(d))$ that are the off-diagonal components of the traceless Hermitean matrix of $su(4)$ Cartan forms
\begin{equation} \label{su4cf}
\Omega_{A}{}^{B}(d)=\left(
\begin{array}{cc}
\Omega_a{}^b& \Omega_a{}^4\\
\Omega_4{}^b& \Omega_4{}^4
\end{array}\right),\quad \Omega_4{}^4=-\Omega_a{}^a.
\end{equation}
Using the isomorphism $SU(4)\sim SO(6)$ it is possible to accommodate $su(4)$ Cartan forms into the $6\times 6$ matrix
\beq
\Omega_{\hat a}{}^{\hat b}(d)=\left(
\begin{array}{cc}
\Omega_{a}{}^b-\delta_a^b\Omega_{c}{}^c & \varepsilon_{acb}\Omega_{4}{}^c\\
-\varepsilon^{acb}\Omega_{c}{}^4 & -\Omega_{b}{}^a+\delta_b^a\Omega_{c}{}^c
\end{array}\right)
\eeq
antisymmetric w.r.t the metric
\begin{equation}
H_{\hat a\hat b}=\left(
\begin{array}{cc}
0 & \delta_a^b\\
\delta^a_b & 0
\end{array}\right)
\end{equation}
following the decomposition of the $D=6$ vector representation as $\mathbf 3\oplus\bar{\mathbf 3}$ of $SU(3)$ \footnote{The metric $H_{\hat a\hat b}$ is the conventional unit $D=6$ metric $\delta^{IJ}$ written in the $\mathbf 3\oplus\bar{\mathbf 3}$ basis. Both bases are connected by the transformation matrices
$$
M^{I\hat a}=\frac12(\tilde\rho^{Ia4},\ \rho^I_{a4}),\quad M^{-1}{}_{\hat aI}=\left(
\begin{array}{c}
\rho^I_{4a}\\ \tilde\rho^{I4a}
\end{array}\right):\quad MM^{-1}=I,
$$
where $\rho^I_{AB}$ and $\tilde\rho^{IAB}$ are $D=6$ chiral $\gamma-$matrices,
such that the components of a $D=6$ vector $O^I$ in these bases can be transformed into one another $O^I=M^{I\hat a}O_{\hat a}$ and $O_{\hat a}=M^{-1}{}_{\hat aI}O^I$. In particular, for the $D=6$ metric we find that $\delta^{IJ}=-2M^{I\hat a}H_{\hat a\hat b}M^{J\hat b}$.}. For the considered choice of the $OSp(4|6)/(SO(1,3)\times U(3))$ supercoset element $su(4)$ Cartan forms are given by the sum of two contributions
\begin{equation}\label{su4cf'}
\Omega_{\hat a}{}^{\hat b}(d)=\Omega_{\mathbf b\hat a}{}^{\hat b}(d)+\Omega_{\mathbf f\hat a}{}^{\hat b}(d)
\end{equation}
coming from bosons and fermions. Bosonic contribution
\beq
\Omega_{\mathbf b\hat a}{}^{\hat b}(d)=iT_{\hat
a}{}^{\hat c}d\bar T_{\hat c}{}^{\hat b}
\eeq
is described by the $su(4)$ Catran form matrix associated with the $SU(4)/U(3)$ coset element
\beq\label{Tmatrix}
T_{\hat a}{}^{\hat b}=\left(
\begin{array}{cc}
T_a{}^b & T_{ab}\\
T^{ab} & T^b{}_a
\end{array}\right)
=\exp\left(
\begin{array}{cc}
0 & i\varepsilon_{acb}z^c\\
-i\varepsilon^{acb}\bar z_c & 0
\end{array}
\right).
\eeq
Explicit expressions for the purely bosonic part of $su(4)$ Cartan forms can be found in \cite{U08}. Fermionic contribution to (\ref{su4cf'})
\beq
\Omega_{\mathbf f\hat a}{}^{\hat b}(d)=(T\Psi(d)\bar T)_{\hat a}{}^{\hat b}
\eeq
is obtained by the $T-$transformation of the matrix
\beq
\Psi_{\hat a}{}^{\hat b}(d)=2(d\theta^\mu_{\hat a}\eta^{\hat b}_\mu-d\theta^{\mu\hat b}\eta_{\mu\hat a}-\eta_{\mu\hat a}\tilde\omega^{\mu\nu}(d)\eta^{\hat b}_\nu),
\eeq
where the Grassmann coordinates have been written as $D=6$ vectors in the $\mathbf 3\oplus\bar{\mathbf 3}$ basis
\beq
\theta^\mu_{\hat a}=\left(
\begin{array}{c}
\theta^\mu_a \\ \bar\theta^{\mu a}
\end{array}\right),\quad \eta_{\mu\hat a}=\left(
\begin{array}{c}
\eta_{\mu a} \\ \bar\eta_{\mu}^{a}
\end{array}\right)
\eeq
and $\theta^{\mu\hat a}=H^{\hat a\hat b}\theta^\mu_{\hat b}$, $\eta_{\mu}^{\hat a}=H^{\hat a\hat b}\eta_{\mu\hat b}$. The $T-$transformed $\mathbf 3\oplus\bar{\mathbf 3}$ vectors will be endowed with hats $\hat\theta^{\mu}_{\hat a}=T_{\hat a}{}^{\hat b}\theta^{\mu}_{\hat b}$, $\hat\theta^{\mu\hat a}=H^{\hat a\hat b}\hat\theta^\mu_{\hat b}$ etc. Using that $H_{\hat a\hat c}(\bar T^T){}^{\hat c}{}_{\hat d}H^{\hat d\hat b}=T_{\hat a}{}^{\hat b}$ for the chosen realization of the matrix $T$ the fermionic part of $su(4)$ Cartan form matrix can be cast into the form
\beq
\Omega_{\mathbf f\hat a}{}^{\hat b}(d)=2(\hat d\theta^\mu_{\hat a}\hat\eta^{\hat b}_\mu-\hat d\theta^{\mu\hat b}\hat\eta_{\mu\hat a}-\hat\eta_{\mu\hat a}\tilde\omega^{\mu\nu}(d)\hat\eta^{\hat b}_\nu).
\eeq

The WZ term of the action (\ref{action}) in the
$\mathbf 3\oplus\bar{\mathbf 3}$ basis can be written in the form
\beq
S_{WZ}=-{\textstyle\frac{i}{8}}\varepsilon^{ij}\mathfrak
J_{\hat a}{}^{\hat b}\int d^2\xi\left(\hat\omega^{\mu\hat
a}_{i}\varepsilon_{\mu\nu}\hat\omega^{\nu}_{j\hat
b}+\hat\chi^{\hat a}_{i\mu}\varepsilon^{\mu\nu}\hat\chi_{j\nu\hat
b}\right),
\eeq
where $\mathfrak J_{\hat a}{}^{\hat b}$ is the
$\mathbb{CP}^3$ Kahler 2-form written in the $\mathbf
3\oplus\bar{\mathbf 3}$ basis\footnote{In conventional $D=6$
vector basis it is given by the expression
$J^{IJ}=\frac{i}{2}(\rho^I_{4a}\tilde\rho^{J4a}-\rho^J_{4a}\tilde\rho^{I4a})$.
It takes simple diagonal form when contracted with the $6d$
rotation generators
$$
J_A{}^B=J^{IJ}\rho^{IJ}{}_A{}^B=\left(
\begin{array}{cc}
-2i & 0\\
0 & 6i
\end{array}\right).
$$
The matrix $J_A{}^B$ can be shown to satisfy the following equation $J_A{}^CJ_C{}^B-4iJ_A{}^B+12\delta_A^B=0$.}
\beq
\mathfrak J_{\hat a}{}^{\hat b}=2i\left(
\begin{array}{cc}
\delta_a^b & 0\\
0 & -\delta_b^a
\end{array}
\right).
\eeq 
It contains the world-sheet projections of fermionic 1-forms
\beq
\hat\omega^\mu_{\hat a}(d)
=e^{-\varphi}T_{\hat a}{}^{\hat b}
\omega^\mu_{\hat b}(d),\quad\omega^\mu_{\hat b}(d)=d\theta^\mu_{\hat b}+\zeta^\mu_{\hat b}(d)
\eeq
related to Poincare supersymmetry generators $(Q^a_\mu, \bar Q_{\mu a})$, and
\begin{equation}
\hat\chi_{\mu\hat a}(d)
=e^{\varphi}T_{\hat a}{}^{\hat b}
\chi_{\mu\hat b}(d),\quad \chi_{\mu\hat a}(d)=d\eta_{\mu\hat a}+2i\eta^{\hat b}_\mu d\theta^\nu_{\hat b}\eta_{\nu\hat a}
-i(\bar\eta\eta)\omega_{\mu\hat a}(d)
\end{equation}
related to conformal supersymmetry generators $(S^{\mu a}, \bar S^{\mu}_a)$.

\setcounter{equation}{0}

\section{$D=3$ $\mathcal N=6$ superconformal symmetry of the $OSp(4|6)/(SO(1,3)\times U(3))$ superstring: general properties and coordinate transformations}

Global $OSp(4|6)$ transformations act on the
$OSp(4|6)/(SO(1,3)\times U(3))$ coset representative from which the left-invariant
Cartan forms (\ref{defcartan}) are constructed in the following way\footnote{Since the supercoset string action is built out of the Cartan forms it is exactly invariant under the global symmetry in distinction with the original Green-Schwarz action \cite{GSorig} that is quasi-invariant because its WZ term cannot be presented as a 2-form in supercurrents. For detailed discussion on that point and the properties of WZ term on $AdS$ backgrounds see, e.g., \cite{Hatsuda}.}
\beq
\mathscr G'H=G\mathscr G,\ G\in OSp(4|6)
\eeq
with $H$ being the
compensating $SO(1,3)\times U(3)$ transformation or passing to
infinitesimal parameters
\beq
\delta\mathscr
G=g\mathscr G-\mathscr Gh,\quad g\in osp(4|6),\quad h\in so(1,3)\oplus u(3).
\eeq
Substituting above relation into (\ref{defcartan}) yields
\beq\label{varcf}
\mathscr C(\delta)=\mathscr G^{-1}\delta\mathscr G=\mathscr
G^{-1}g\mathscr G-h.
\eeq

Consider the $D=3$ $\mathcal N=6$ superconformal algebra valued transformation parameter 
\beq
\begin{array}{rl}
g=&a^mP_m+b_mK^m+fD+\frac12l^{mn}M_{mn}+y^aV_a{}^4+\bar
y_aV_4{}^a+w_a{}^bV_b{}^a+w_4{}^4V_4{}^4\\
+&\varepsilon^\mu_a Q^a_\mu+\bar\varepsilon^{\mu a}\bar Q_{\mu
a}+\xi_{\mu a}S^{\mu a}+\bar\xi^a_\mu\bar S^\mu_a.
\end{array}
\eeq
It includes the parameters of $D=3$ Minkowski space-time translations $a^m$, conformal boosts $b_m$, dilatations $f$ and Lorentz rotations $l^{mn}$, as well as anticommuting parameters of $D=3$ $\mathcal N=6$ Poincare supersymmetry $(\varepsilon^\mu_a, \bar\varepsilon^{\mu a})$ and conformal supersymmetry $(\xi_{\mu a}, \bar\xi^a_\mu)$ supplemented by the $SU(4)$ $R-$symmetry parameters $(w_a{}^b, y^a, \bar y_a)$. Then the substitution of $OSp(4|6)/(SO(1,3)\times U(3))$ coset representative (\ref{scosetelement}) into (\ref{varcf}) yields
\beq
\begin{array}{rl}
\mathscr C(\delta)=&(\hat\omega^m(\delta)-\hat b^m)P_{m}+(\hat c^m(\delta)+\hat b^m)K_m+\Delta(\delta)D+(G^{mn}(\delta)+\frac12\hat l^{mn})M_{mn}\\
+&\Omega_4{}^a(\delta)V_a{}^4+\Omega_a{}^4(\delta)V_4{}^a+(\Omega_a{}^b(\delta)+\hat w_a{}^b)V_b{}^a+(\Omega_4{}^4(\delta)+\hat w_4{}^4)V_4{}^4\\
+&\hat\omega^\mu_a(\delta)Q^a_\mu+\hat{\bar\omega}{}^{\mu
a}(\delta)Q_{\mu a}+\hat\chi_{\mu a}(\delta)S^{\mu
a}+\hat{\bar\chi}{}^a_\mu(\delta)\bar S^\mu_a.
\end{array}
\eeq
The quantities that cannot be accommodated into the individual
Cartan form variations like, e.g.
$\hat\omega^m(\delta)=i_\delta\hat\omega^m(d)$ represent
parameters of the compensating transformations. In particular, the
vector
\beq\label{compso13so12} \hat
b^m=e^{2\varphi}A^{-1}b^m(\theta),\quad
A=1-e^{4\varphi}(\bar\eta\eta)^2,
\eeq
where
\beq
b^m(\theta)=b^m-i\left[(\xi_a(\theta)\tilde\sigma^m\bar\eta^a)+(\bar\xi^a(\theta)\tilde\sigma^m\eta_a)\right],\quad\xi_{\mu
a}(\theta)=\xi_{\mu a}+b_{\mu\nu}\theta^\nu_a,
\eeq
is the parameter of
$SO(1,3)/SO(1,2)$ transformations, while the antisymmetric tensor
\beq\label{hatl} \hat
l^{mn}=l^{mn}(\theta)+ie^{2\varphi}\left[(\eta_a\hat{\tilde
b}\sigma^{mn}\bar\eta^a)+(\bar\eta^a\hat{\tilde
b}\sigma^{mn}\eta_a)\right]
\eeq
with
\beq
l^{mn}(\theta)=l^{mn}\!+\!2(b^mx^n-b^nx^m)\!+\!2i\!\left[(\theta_a\sigma^{mn}\bar\xi^a)+(\bar\theta^a\sigma^{mn}\xi_a)\right]\!+\!i\!\left[(\theta_a\sigma^{mn}b\bar\theta^a)+(\bar\theta^a\sigma^{mn}b\theta_a)
\right]
\eeq
describes compensating $SO(1,2)$ Lorentz rotations. The parameters of
compensating $U(3)$ rotations
\beq\label{compu3} \hat
w_a{}^b=\widetilde
w_a{}^b+{\textstyle\frac{i(1-\cos{|z|})}{|z|\sin{|z|}}}(\bar
z_a\widetilde y^b-\widetilde{\bar
y}_az^b)+{\textstyle\frac{i(1-\cos{|z|})^2}{2|z|^3\cos{|z|}\sin{|z|}}}((\widetilde
y\bar z)-(z\widetilde{\bar y}))\bar z_az^b,\quad |z|^2=z^a\bar z_a
\eeq
have been presented in the form exhibiting explicit
dependence on the entries
\beq
\begin{array}{c}
\widetilde w_a{}^b=w_a{}^b(\theta)-e^{2\varphi}\left[2(\eta_a\hat{\tilde
b}\bar\eta^b)-\delta_a^b(\eta_c\hat{\tilde b}\bar\eta^c)\right],\\[0.2cm]
\widetilde y^a=y^a(\theta)+e^{2\varphi}\varepsilon^{abc}(\eta_b\hat{\tilde
b}\eta_c),\quad\widetilde{\bar y}_a=\bar
y_a(\theta)-e^{2\varphi}\varepsilon_{abc}(\bar\eta^b\hat{\tilde
b}\bar\eta^c),
\end{array}
\eeq
where
\beq
\begin{array}{c}
w_a{}^b(\theta)=w_a{}^b-2(\xi_{\mu a}\bar\theta^{\mu
b}+\theta^\mu_a\bar\xi^b_\mu)+\delta_a^b(\xi_{\mu
c}\bar\theta^{\mu
c}+\theta^\mu_c\bar\xi^c_\mu)-2(\theta_ab\bar\theta^b)+\delta_a^b(\theta_cb\bar\theta^c),\\[0.2cm]
y^a(\theta)=y^a+\varepsilon^{abc}\left(2\xi_{\mu
b}\theta^\mu_c+(\theta_bb\theta_c)\right),\quad \bar y_a(\theta)=\bar
y_a-\varepsilon_{abc}\left(2\bar\xi_{\mu}^{b}\bar\theta^{\mu
c}+(\bar\theta^bb\bar\theta^c)\right),
\end{array}
\eeq
of the $SU(4)$ matrix
\beq\label{Wmatrix}
\widetilde W_{\hat a}{}^{\hat b}=\left(
\begin{array}{cc}
\widetilde w_a{}^b-\delta_a^b\widetilde w_c{}^c & \varepsilon_{acb}\widetilde
y^c\\[0.2cm]
-\varepsilon^{acb}\widetilde{\bar y}_c & -\widetilde w_b{}^a+\delta_b^a\widetilde
w_c{}^c
\end{array}\right)
\eeq
that, as will be shown below, enters the transformation laws (\ref{su4u3matrixtrans}) of the $SU(4)/U(3)$ coset element (\ref{Tmatrix}) under the $D=3$ $\mathcal N=6$ superconformal symmetry.

$D=3$ $\mathcal N=6$ superconformal transformations of the $OSp(4|6)/(SO(1,3)\times U(3))$ superspace coordinates that parametrize (\ref{scosetelement}) include also the
contributions proportional to the parameters of the compensating
transformations (\ref{compso13so12}), (\ref{hatl}) and (\ref{compu3}). $D=3$ $\mathcal N=6$ Boundary superspace coordinates obey the following transformation rules\footnote{Observe vanishing of the terms proportional to $SO(1,3)/SO(1,2)$ rotation parameters $\hat b$ in the boundary limit $\varphi\to-\infty$.}
\beq
\begin{array}{rl}
\delta x^m=& a^m+l^m{}_nx^n+2fx^m+b^m(x^2+(\bar\theta\theta)^2)-2x^mb_nx^n\\
 -&i\left[(\varepsilon_a\sigma^m\bar\theta^a)+(\bar\varepsilon^a\sigma^m\theta_a)\right]-i\left[(\xi_a\hat{\tilde x}\sigma^m\bar\theta^a)+(\bar\xi^a\hat{\tilde x}\sigma^m\theta_a)\right]\\
+&e^{2\varphi}\left\{\hat
b^m+i\left[(\eta_a\hat{\tilde
b}\sigma^m\bar\theta^a)+(\bar\eta^a\hat{\tilde b}\sigma^m\theta_a)\right]\right\},
\end{array}
\eeq
\beq
\begin{array}{rl}
\delta\theta^\mu_a=&\varepsilon^\mu_a+\frac14l^{mn}\theta^\nu_a\sigma_{mn\nu}{}^\mu+f\theta^\mu_a+iw_b{}^b\theta^\mu_a-iw_a{}^b\theta^\mu_b-i\varepsilon_{abc}y^b\bar\theta^{\mu c}+\hat{\tilde
x}{}^{\mu\nu}b_{\nu\lambda}\theta^\lambda_a\\
+&\hat{\tilde x}{}^{\mu\nu}\xi_{\nu a}-2i(\theta^\mu_b\bar\xi^b_\nu
+\bar\theta^{\mu b}\xi_{\nu b})\theta^\nu_a+e^{2\varphi}\hat{\tilde b}{}^{\mu\nu}\eta_{\nu a}
\end{array}
\eeq
and c.c., where $\hat{\tilde x}{}^{\mu\nu}=\tilde x^{\mu\nu}-i\varepsilon^{\mu\nu}(\bar\theta\theta)$,
while that for the coordinate $\varphi$ related to the $AdS_4$ space bulk direction reads
\beq
\delta\varphi=f(\theta)=f-b_mx^m+i(\xi_{\mu
a}\bar\theta^{\mu a}+\bar\xi^a_\mu\theta^\mu_a).
\eeq
Transformation properties of the $\mathbb{CP}^3$ complex coordinates
\beq
\begin{array}{rl}
\delta z^a=&iz^b\widetilde w_b{}^a+i\widetilde w_b{}^bz^a+\frac{|z|\cos{|z|}}{\sin{|z|}}\widetilde y^a+\frac{1}{2|z|^2}\left(1-\frac{|z|}{\cos{|z|}\sin{|z|}}\right)(\widetilde y\bar z)z^a\\[0.2cm]
+&\frac{1}{2|z|^2}\left[1+|z|(\tan{|z|}-\cot{|z|})\right](z\widetilde{\bar y})z^a
\end{array}
\eeq
can be summarized in the form of $SU(4)/U(3)$ coset representative (\ref{Tmatrix}) transformations
\beq\label{su4u3matrixtrans}
\delta T_{\hat a}{}^{\hat b}=i(T_{\hat
a}{}^{\hat c}\widetilde W_{\hat c}{}^{\hat b}-\widehat W_{\hat a}{}^{\hat c}T_{\hat c}{}^{\hat b}),\quad \delta\bar T_{\hat
a}{}^{\hat b}=-i(\widetilde W_{\hat a}{}^{\hat c}\bar T_{\hat c}{}^{\hat b}-\bar T_{\hat
a}{}^{\hat c}\widehat W_{\hat c}{}^{\hat b}),
\eeq
where the $SU(4)$ matrix $\widetilde W_{\hat a}{}^{\hat b}$ has been introduced in (\ref{Wmatrix}) and
\beq
\widehat W_{\hat a}{}^{\hat b}=\left(
\begin{array}{cc}
\hat w_a{}^b-\delta_a^b\hat w_c{}^c & 0\\[0.2cm]
0 & -\hat w_b{}^a+\delta_b^a\hat w_c{}^c
\end{array}\right)
\eeq
represents $U(3)$ compensating rotation matrix.
Finally Grassmann coordinates associated with the conformal supersymmetry generators transform as follows
\beq
\begin{array}{rl}
\delta\eta_{\mu a}=&\xi_{\mu
a}(\theta)-\frac14l^{mn}(\theta)\sigma_{mn\mu}{}^\nu\eta_{\nu
a}-f(\theta)\eta_{\mu a}+iw_b{}^b(\theta)\eta_{\mu
a}-iw_a{}^b(\theta)\eta_{\mu
b}-i\varepsilon_{abc}y^b(\theta)\bar\eta^c_\mu\\[0.2cm]
+&2ie^{2\varphi}(\bar\eta\eta)\varepsilon_{\mu\lambda}\hat{\tilde b}{}^{\lambda\nu}\eta_{\nu a}\\[0.2cm]
=&-\frac14l^{mn}\sigma_{mn\mu}{}^\nu\eta_{\nu a}-f\eta_{\mu a}+iw_b{}^b\eta_{\mu a}-iw_a{}^b\eta_{\mu b}-i\varepsilon_{abc}y^b\bar\eta^c_\mu+b_{\mu\nu}\theta^\nu_a-\eta_{\nu a}\hat{\tilde x}{}^{\nu\lambda}b_{\lambda\mu}\\[0.2cm]
+&2i\left[(\theta_ab\theta_b)\bar\eta^b_\mu+(\theta_ab\bar\theta^b)\eta_{\mu b}\right]+\xi_{\mu a}-2i(\bar\xi^b_\mu\theta^\nu_b+\xi_{\mu b}\bar\theta^{\nu b})\eta_{\nu a}\\[0.2cm]
-&2i(\eta_{\mu b}\bar\xi^b_\nu+\bar\eta^b_\mu\xi_{\nu b})\theta^\nu_a+2i\xi_{\nu a}(\theta^\nu_b\bar\eta^b_\mu+\bar\theta^{\nu b}\eta_{\mu b})+2ie^{2\varphi}(\bar\eta\eta)\varepsilon_{\mu\lambda}\hat{\tilde b}{}^{\lambda\nu}\eta_{\nu a}
\end{array}
\eeq
and c.c.

Cartan forms associated with the $osp(4|6)/(so(1,3)\times u(3))$ supercoset generators are left-invariant under the above derived global transformations up to the compensating ones associated with the stability group generators. Corresponding Cartan forms in their turn transform in a connection-type way. In particular, bosonic 1-forms that are identified with the $AdS_4$ part of the supervielbein in general transform as
\beq\label{genvaradscf}
\delta\hat\omega^m(d)+\delta\hat c^m(d)=\hat l^{mn}(\hat\omega_{n}(d)+\hat c_{n}(d))+4\hat b^m\Delta(d),\quad \delta\Delta(d)=-\hat
b_m(\hat\omega^{m}(d)+\hat c^{m}(d)),
\eeq
while, e.g. $so(1,2)$ Cartan forms in the spinor realization $G^{\mu\nu}(d)=\varepsilon^{\mu\lambda}\sigma_{mn\lambda}{}^\nu G^{mn}(d)$ obey the following rule
\beq
\delta G^{\mu\nu}(d)={\textstyle\frac14}(G^{\mu\lambda}(d)\hat l_\lambda{}^\nu+G^{\nu\lambda}(d)\hat l_\lambda{}^\mu)+\hat b^\mu{}_\lambda(\hat{\tilde\omega}(d)-\hat{\tilde c}(d))^{\lambda\nu}+\hat b^\nu{}_\lambda(\hat{\tilde\omega}(d)-\hat{\tilde c}(d))^{\lambda\mu}-{\textstyle\frac12}d\hat l^{\mu\nu}.
\eeq
$su(4)$ Cartan forms are $OSp(4|6)$ left-invariant up to the $U(3)$ gauge transformation
\beq\label{genvarsu4cf}
\delta\Omega_{\hat a}{}^{\hat b}(d)=i(\Omega_{\hat a}{}^{\hat c}(d)\widehat W_{\hat c}{}^{\hat b}-\widehat W_{\hat a}{}^{\hat c}\Omega_{\hat c}{}^{\hat b}(d))-d\widehat W_{\hat a}{}^{\hat b},
\eeq
from where we infer that the $su(4)/u(3)$ 1-forms identified with the $\mathbb{CP}^3$ part of the supervielbein transform as
\beq\label{genvarcp3cf}
\delta\Omega_a{}^4(d)=i\left(\hat w_b{}^b\Omega_a{}^4(d)-\hat w_a{}^b\Omega_b{}^4(d)\right),\quad\delta\Omega_4{}^a(d)=-i\left(\hat w_b{}^b\Omega_4{}^a(d)-\Omega_4{}^b(d)\hat w_b{}^a\right)
\eeq
and $u(3)$ 1-forms exhibit connection-type transformation properties
\beq
\delta\Omega_a{}^b(d)=i(\Omega_a{}^c(d)\hat w_c{}^b-\hat w_a{}^c\Omega_c{}^b(d))-d\hat w_a{}^b.
\eeq
Cartan forms that are identified with the supervielbein fermionic components transform in the following way
\beq\label{genvarspoincarecf}
\delta\hat\omega_{\hat a}^\nu(d)=\frac14\hat\omega^\lambda_{\hat a}(d)\hat l_{\lambda}{}^\nu+\hat{\tilde b}{}^{\nu\lambda}\hat\chi_{\lambda\hat a}(d)-i\widehat W_{\hat a}{}^{\hat b}\hat\omega_{\hat b}^\nu(d)
\eeq
and
\beq\label{genvarsconfcf}
\delta\hat\chi_{\nu\hat  a}(d)=-\frac14\hat l_{\nu}{}^\lambda\hat\chi_{\lambda\hat a}(d)+\hat b_{\nu\lambda}\hat\omega^\lambda_{\hat a}(d)-i\widehat W_{\hat a}{}^{\hat b}\hat\chi_{\nu\hat b}(d).
\eeq
For individual transformations from the $D=3$ $\mathcal N=6$ superconformal symmetry to be discussed below these expressions simplify by properly restricting the parameters of compensating transformations that will be indicated by the vertical line.

\setcounter{equation}{0}

\section{$D=3$ $\mathcal N=6$ superconformal symmetry of the $OSp(4|6)/(SO(1,3)\times U(3))$ superstring: Noether currents}

Noether current densities corresponding to the $D=3$ $\mathcal N=6$ superconformal invariance of the superstring action (\ref{action}) can be formally presented as the sum
\beq
\mathcal J^{i}_{\Sigma}(\tau,\sigma)=\mathcal J_{AdS}{}^{i}_{\Sigma}+\mathcal J_{CP}{}^{i}_{\Sigma}+\mathcal J_{WZ}{}^{i}_{\Sigma},
\eeq
where $\Sigma$ is a transformation parameter\footnote{We assume the right derivative for
fermions.}, of contributions of the  $AdS_4$
\beq
\mathcal J_{AdS}{}^{i}_{\Sigma}=-\sqrt{-g}g^{ij}\left({\textstyle\frac{1}{4}}(\hat\omega_{jm}+\hat
c_{jm}){\textstyle\frac{\partial}{\partial\Sigma}}(\hat\omega^m(\delta_\Sigma)+\hat
c^m(\delta_\Sigma))+\Delta_j{\textstyle\frac{\partial}{\partial\Sigma}}\Delta(\delta_\Sigma)\right)
\eeq
and $\mathbb{CP}^3$ parts of the
kinetic term
\beq
\mathcal J_{CP}{}^{i}_{\Sigma}=-{\textstyle\frac12}\sqrt{-g}g^{ij}\left(\Omega_{ja}{}^4{\textstyle\frac{\partial}{\partial\Sigma}}\Omega_4{}^a(\delta_\Sigma)+\Omega_{j4}{}^a{\textstyle\frac{\partial}{\partial\Sigma}}\Omega_a{}^4(\delta_\Sigma)\right),
\eeq
as well as that of the Wess-Zumino term
\beq
\mathcal J_{WZ}{}^{i}_{\Sigma}={\textstyle\frac{i}{4}}\varepsilon^{ij}\mathfrak J_{\hat a}{}^{\hat b}\left(\hat{\omega}^{\mu\hat a}_j\varepsilon_{\mu\nu}{\textstyle\frac{\partial}{\partial\Sigma}}\hat\omega^\nu_{\hat b}(\delta_\Sigma)+\hat{\chi}^{\hat a}_{j\mu}\varepsilon^{\mu\nu}{\textstyle\frac{\partial}{\partial\Sigma}}\hat\chi_{\nu\hat b}(\delta_\Sigma)\right).
\eeq
Below we specialize to discussion of the individual transformations from the $D=3$ $\mathcal N=6$ superconformal symmetry and present corresponding expressions for the Noether currents.

\subsection{Noether currents associated with $D=3$ conformal symmetry}

\subsubsection{Space-time translations}

$osp(4|6)$ Cartan forms are obviously invariant under the global translations of $D=3$ Minkowski boundary coordinates. Their contributions to the current density are hence related to the coordinate dependence of transformation parameter. In particular, Eq.(\ref{genvaradscf}) representing variation of the Cartan forms identified with the $AdS$ part of the supervielbein, when restricted to the boundary space-time translations acquires the form
\beq
\delta_{a}\hat\omega^m(d)+\delta_{a}\hat c^m(d)=j^{m}{}_nda^{n},\quad \delta_a\Delta(d)=0,
\eeq
where the current contribution tensor equals
\beq \label{transadscf}
j^{m}{}_n=\frac{\partial(\hat\omega^m(\delta_a)+\hat c^m(\delta_a))}{\partial a^n}=e^{-2\varphi}A\delta^m_n.
\eeq
$su(4)$ Cartan forms are also invariant under $3d$ translations
\beq
\delta_a\Omega_{\hat a}{}^{\hat b}(d)=J_{\hat a}{}^{\hat b}{}_mda^m
\eeq
modulo the current contribution matrix
\beq
J_{\hat a}{}^{\hat b}{}_m=\frac{\partial}{\partial a^m}\Omega_{\hat a}{}^{\hat b}(\delta_a)=\left(
\begin{array}{cc}
j_a{}^{b}{}_m & j_{abm}\\[0.2cm]
-\bar j^{ab}{}_m & -j_b{}^{a}{}_m
\end{array}\right)=-2(\hat\eta_{\hat a}\sigma_m\hat\eta^{\hat b}).
\eeq
As a result variation of the $\mathbb{CP}^3$ components of the supervielbein acquires the form
\beq \label{transcp3viel}
\delta_a\Omega_{a}{}^4(d)=-{}^*\!\bar j_{am}da^m,\quad
\delta_a\Omega_{4}{}^a(d)=-{}^*\! j^a{}_mda^m,
\eeq
where we have adopted the following definition of the $SU(3)$ vector dual to a rank 2 tensor that can also carry other indices $\Sigma$
\beq
({}^*\! j^a)^\Sigma=\frac12\varepsilon^{abc}j_{bc}{}^\Sigma,\quad({}^*\! j_a)^\Sigma=\frac12\varepsilon_{abc}j^{bc\Sigma}
\eeq
or simply ${}^*\! j^{a\Sigma}$ and ${}^*\! j_a{}^\Sigma$ if $\Sigma$ does not contain $SU(3)$ indices. 
Variation of the fermionic supervielbein components associated with the Poincare supersymmetry under localized boundary space-time translations reads
\beq
\delta_a\hat\omega^\mu_{\hat a}(d)=j^\mu_{\hat am}da^m
\eeq
with the current contribution
\beq \label{transspoincarecf}
j^\mu_{\hat am}=\frac{\partial\hat\omega^\mu_{\hat a}(\delta_a)}{\partial a^m}=-e^{-\varphi}\tilde\sigma^{\mu\nu}_m\hat\eta_{\nu\hat a}.
\eeq
Then for the variation of remaining fermionic supervielbein components associated with the conformal supersymmetry we obtain
\beq
\delta_a\hat\chi_{\mu\hat a}(d)=J_{\mu\hat am}da^m,
\eeq
where
\beq \label{transsconfcf}
J_{\mu\hat am}=\frac{\partial\hat\chi_{\mu\hat a}(\delta_a)}{\partial a^m}=-ie^{2\varphi}(\bar\eta\eta)\varepsilon_{\mu\nu}j^\nu_{\hat am}
\eeq
is the current contribution.

The current density related to the superstring action
(\ref{action}) invariance under $D=3$ Minkowski space-time
translations has the form
\beq
\mathcal
J{}^{i}_m(\tau,\sigma)=\mathcal J_{AdS}{}^{i}_{m}+\mathcal
J_{CP}{}^{i}_{m}+\mathcal J_{WZ}{}^{i}_{m}.
\eeq
The $AdS$ part of
the current density
\beq
\mathcal
J_{AdS}{}^{i}_{m}=-\frac{1}{4}\sqrt{-g}g^{ij}(\hat\omega_{jn}+\hat
c_{jn})j^{n}{}_{m}
\eeq
is determined by the current contribution
(\ref{transadscf}), the $\mathbb{CP}^3$ part
\beq
\mathcal
J_{CP}{}^{i}_{m}=\frac12\sqrt{-g}g^{ij}\left(\Omega_{ja}{}^4\;{}^*\!
j^a{}_{m}+\Omega_{j4}{}^a\;{}^*\!\bar j_{am}\right)
\eeq
is
contributed by Eq.(\ref{transcp3viel}), and the current
contributions (\ref{transspoincarecf}), (\ref{transsconfcf})
determine the WZ part of the current density
\beq
\mathcal
J_{WZ}{}^{i}_{m}=\frac{i}{4}\varepsilon^{ij}\mathfrak J_{\hat
a}{}^{\hat b}\left(\hat\omega^{\mu\hat
a}_{j}\varepsilon_{\mu\nu}j^{\nu}_{\hat
bm}+\hat\chi_{j\mu}^{\hat a}\varepsilon^{\mu\nu} J_{\nu\hat
bm}\right).
\eeq

\subsubsection{Conformal boosts}

Transformation properties of the Cartan forms that enter the $AdS$ part of the $OSp(4|6)/(SO(1,3)\times U(3))$ superstring action (\ref{action}) under local conformal boosts follow from the expressions (\ref{genvaradscf}) appropriately restricted
\beq \label{adsconfcur}
\begin{array}{rl}
\delta_{b}\,\hat\omega^m(d)+\delta_{b}\hat c^m(d) &=j^{mn}db_{n}+(\hat l|_b)^{mn}(\hat\omega_{n}(d)+\hat c_{n}(d))+4(\hat b|_b)^m\Delta(d),\\
\delta_b\Delta(d)&=j^{m}db_{m}-(\hat b|_b)^m(\hat\omega_{m}(d)+\hat c_{m}(d))
\end{array}
\eeq
modulo the current contributions
\beq
\begin{array}{rl}
j^{mn}=&\frac{\partial(\hat\omega^m(\delta_b)+\hat c^m(\delta_b))}{\partial b_n}=e^{-2\varphi}A\!\left\{\!(x^2\!+\!(\bar\theta\theta)^2)\eta^{mn}\!-\!2x^mx^n\!+\! i\!\left[(\theta_a\sigma^m\hat{\tilde x}\sigma^n\bar\theta^a)\!+\!(\bar\theta^a\sigma^m\hat{\tilde x}\sigma^n\theta_a)\right]\!\right\}\\[0.2cm]
+&A\frac{\partial\hat b^m}{\partial b_n}+ie^{2\varphi}\left[(\eta_a\hat{\tilde x}\sigma^n\tilde\sigma^m\bar\eta^a)+(\bar\eta^a\hat{\tilde x}\sigma^n\tilde\sigma^m\eta_a)+(\eta_a\tilde\sigma^m\Lambda_-\sigma^n\bar\theta^a)+(\bar\eta^a\tilde\sigma^m\Lambda_-\sigma^n\theta_a)\right]\\[0.2cm]
+&2e^{2\varphi}(\bar\eta\eta)\left[(\eta_a\sigma^m\hat{\tilde x}\sigma^n\bar\theta^a)+(\bar\eta^a\sigma^m\hat{\tilde x}\sigma^n\theta_a)\right],
\end{array}
\eeq
where we have introduced the following $3d$ spin-tensors
$\Lambda_{\pm\mu}{}^\nu=\delta_\mu^\nu\pm2i(\bar\eta^a_\mu\theta^\nu_a+\eta_{\mu
a}\bar\theta^{\nu a})$ that will appear to be useful below, and
\beq
j^m=\frac{\partial\Delta(\delta_b)}{\partial b_m}=-x^m-i\left[(\eta_a\hat{\tilde x}\sigma^m\bar\theta^a)+(\bar\eta^a\hat{\tilde x}\sigma^m\theta_a)\right].
\eeq
Variation of the $su(4)$ Cartan forms is obtained by specializing to the conformal boost parameter dependence in Eq.(\ref{genvarsu4cf})
\beq \label{bvarOmega}
\delta_b\Omega_{\hat a}{}^{\hat b}(d)=\hat J_{\hat a}{}^{\hat bm}db_m+i\left(\Omega_{\hat a}{}^{\hat c}(d)(\widehat W|_b)_{\hat c}{}^{\hat b}-(\widehat W|_b)_{\hat a}{}^{\hat b}\Omega_{\hat b}{}^{\hat c}(d)\right)-d(\widehat W|_b)_{\hat a}{}^{\hat b}.
\eeq
The current contribution matrix
\beq\label{confsu4ccm}
\hat J_{\hat a}{}^{\hat bm}=\frac{\partial}{\partial b_m}\Omega_{\hat a}{}^{\hat b}(\delta_b)=\left(
\begin{array}{cc}
\hat j_a{}^{bm} & \hat j_{ab}{}^m\\[0.2cm]
-\hat{\bar j}{}^{abm} & -\hat j_b{}^{am}
\end{array}\right)
\eeq
is obtained by $T-$transforming the matrix
\beq
\begin{array}{rl}
J_{\hat a}{}^{\hat bm}=&\textstyle{\frac{\partial}{\partial b_{m}}}((\widetilde W|_b)_{\hat a}{}^{\hat b}+\Psi_{\hat a}{}^{\hat b}(\delta_b))=-2\left[(\theta_{\hat a}\sigma^m\theta^{\hat b})+(x^2+(\bar\theta\theta)^2)(\eta_{\hat a}\tilde\sigma^m\eta^{\hat b})\right.\\[0.2cm]
-&\left.2x^m(\eta_{\hat a}\tilde x\eta^{\hat b})-(\eta_{\hat a}\hat{\tilde x}\sigma^mZ^{\hat b})+(\eta^{\hat b}\hat{\tilde x}\sigma^mZ_{\hat a})\right],
\end{array}
\eeq
where $Z^\mu_{\hat a}=\theta^\mu_{\hat a}-i(\bar\theta\theta)\eta^\mu_{\hat a}$. The final form of the current contribution matrix (\ref{confsu4ccm}) is
\beq
\begin{array}{rl}
\hat J_{\hat a}{}^{\hat bm}=&(TJ^m\bar T)_{\hat a}{}^{\hat b}=-2\left[(\hat\theta_{\hat a}\sigma^m\hat\theta^{\hat b})+(x^2+(\bar\theta\theta)^2)(\hat\eta_{\hat a}\tilde\sigma^m\hat\eta^{\hat b})\right.\\[0.2cm]
-&\left.2x^m(\hat\eta_{\hat a}\tilde x\hat\eta^{\hat b})-(\hat\eta_{\hat a}\hat{\tilde x}\sigma^m\hat Z^{\hat b})+(\hat\eta^{\hat b}\hat{\tilde x}\sigma^m\hat Z_{\hat a})\right].
\end{array}
\eeq
Thus variation of the supervielbein components tangent to the $\mathbb{CP}^3$ manifold is brought to the form
\beq \label{cpconfcur}
\delta_b\Omega_{a}{}^4(d)=-{}^*\!\hat{\bar j}_{a}{}^mdb_m+i(\hat w|_b)_b{}^b\Omega_a{}^4(d)-i(\hat w|_b)_a{}^b\Omega_b{}^4(d)
\eeq 
and c.c. 
The expressions for the variation of fermionic 1-forms follow from (\ref{genvarspoincarecf}) and (\ref{genvarsconfcf}). Namely, for Cartan forms related to Poincare supersymmetry one derives that
\beq \label{spoincareconfcur}
\delta_b\hat\omega^\mu_{\hat a}(d)=j^{\mu m}_{\hat a}db_m+\frac14\hat\omega^\nu_{\hat a}(d)(\hat l|_b)_{\nu}{}^\mu+(\hat{\tilde b}|_b)^{\mu\nu}\hat\chi_{\nu\hat a}(d)-i(\widehat W|_b)_{\hat a}{}^{\hat b}\hat\omega_{\hat b}^\mu(d),
\eeq
where the current contribution reads
\beq
j^{\mu m}_{\hat a}=\frac{\partial\hat\omega^\mu_{\hat a}(\delta_b)}{\partial b_m}=e^{-\varphi}\!\left[\hat{\tilde x}{}^{\mu\nu}\sigma^m_{\nu\lambda}\hat Z^\lambda_{\hat a}\!-\!(x^2+(\bar\theta\theta)^2)\tilde\sigma^{m\mu\nu}\hat\eta_{\nu\hat a}\!+\!2x^m\tilde x^{\mu\nu}\hat\eta_{\nu\hat a}\!+\!i(\bar\theta\theta)\varepsilon^{\mu\nu}\hat\eta_{\rho\hat a}\hat{\tilde x}{}^{\rho\lambda}\sigma^m_{\lambda\nu}\right]\!.
\eeq
Correspondingly for Cartan forms related to conformal supersymmetry we find
\beq \label{sconfconfcur}
\delta_b\hat\chi_{\mu\hat a}(d)=J_{\mu\hat a}{}^mdb_m-
\frac14(\hat l|_b)_{\mu}{}^\nu\hat\chi_{\nu\hat a}(d)+(\hat b|_b)_{\mu\nu}\hat\omega^\nu_{\hat a}(d)-i(\widehat W|_b)_{\hat a}{}^{\hat b}\hat\chi_{\mu\hat b}(d)
\eeq
with the current contribution
\beq
J_{\mu\hat a}{}^m=\frac{\partial\hat\chi_{\mu\hat a}(\delta_b)}{\partial b_m}=e^{\varphi}\left(\Lambda_{-\mu}{}^\nu\sigma^m_{\nu\lambda}\hat\theta^\lambda_{\hat a}-\hat\eta_{\nu\hat a}\hat{\tilde x}{}^{\nu\lambda}\sigma^m_{\lambda\rho}\Lambda_{-\mu}{}^\rho-ie^{\varphi}(\bar\eta\eta)\varepsilon_{\mu\nu}j^\nu_{\hat a}{}^m\right).
\eeq

The substitution of Eqs.(\ref{adsconfcur}), (\ref{cpconfcur}), (\ref{spoincareconfcur}) and (\ref{sconfconfcur}) into the superstring action variation under $D=3$ conformal boosts yields the current density
\beq
\mathcal J^{im}(\tau,\sigma)=\mathcal J_{AdS}{}^{im}+\mathcal J_{CP}{}^{im}+\mathcal J_{WZ}{}^{im},
\eeq
where
\beq
\begin{array}{c}
\mathcal J_{AdS}{}^{im}=-\sqrt{-g}g^{ij}\left(\frac{1}{4}(\hat\omega_{jn}+\hat
c_{jn})j^{nm}+\Delta_jj^{m}\right),\\[0.2cm]
\mathcal J_{CP}{}^{im}=\frac12\sqrt{-g}g^{ij}\left(\Omega_{ja}{}^4\;{}^*\!\hat j^{am}+\Omega_{j4}{}^a\;{}^*\!\hat{\bar j}_a{}^{m}\right),\\[0.2cm]
\mathcal J_{WZ}{}^{im}=\frac{i}{4}\varepsilon^{ij}\mathfrak J_{\hat a}{}^{\hat b}\left(\hat{\omega}^{\mu\hat a}_j\varepsilon_{\mu\nu}j^{\nu m}_{\hat b}+\hat{\chi}^{\hat a}_{j\mu}\varepsilon^{\mu\nu}J_{\nu\hat b}{}^m\right).
\end{array}
\eeq

\subsubsection{Dilatations}

$osp(4|6)/(so(1,3)\times u(3))$ Cartan forms identified with the $(10|24)-$supervielbein components are invariant under the global scale transformations due to the presence of appropriate exponents of the $AdS_4$ bulk coordinate $\varphi$. Hence their variation under coordinate-dependent scale transformations is determined by the current contributions. In particular, for the components of supervielbein tangent to the $AdS_4$ space we obtain that
\beq
\delta_f\,\hat\omega^m(d)+\delta_f\hat c^m(d)=j^mdf,\quad\delta_f\Delta(d)=jdf,
\eeq
where
\beq \label{diladscf}
\begin{array}{rl}
j^m=&\frac{\partial(\hat\omega^m(\delta_f)+\hat c^m(\delta_f))}{\partial f}=2e^{-2\varphi}Ax^m+2e^{2\varphi}(\bar\eta\eta)\left[(\eta_a\sigma^m\bar\theta^a)+(\bar\eta^a\sigma^m\theta_a)\right],\\[0.2cm]
j=&\frac{\partial\Delta(\delta_f)}{\partial f}=1+i(\theta^\mu_a\bar\eta^a_\mu+\bar\theta^{\mu a}\eta_{\mu a}).
\end{array}
\eeq
$su(4)$ Cartan forms are also scale-invariant
\beq
\delta_f\Omega_{\hat a}{}^{\hat b}=J_{\hat a}{}^{\hat b}df
\eeq
modulo the current contribution matrix 
\beq\label{su4ccmdilatations}
J_{\hat a}{}^{\hat b}=\frac{\partial}{\partial f}\Omega_{\hat a}{}^{\hat b}(\delta_f)=\left(
\begin{array}{cc}
j_a{}^{b} & j_{ab}\\[0.2cm]
-\bar j^{ab} & -j_b{}^{a}
\end{array}\right)=2\left(\hat\Theta^\mu_{\hat a}\hat\eta^{\hat b}_\mu-\hat\Theta^{\mu\hat b}\hat\eta_{\mu\hat a}\right),
\eeq
where $ \Theta^\mu_{\hat b}=\theta^\mu_{\hat b}-\eta_{\nu\hat
b}\hat{\tilde x}{}^{\nu\mu}$. So that the variation of
$\mathbb{CP}^3$ part of the supervielbein is governed by the
appropriate components of the matrix
(\ref{su4ccmdilatations})
\beq \label{dilcp3viel}
\delta_f\Omega_a{}^4(d)=-{}^*\!\bar
j_adf,\quad\delta_f\Omega_4{}^a(d)=-{}^*\! j^adf.
\eeq
Fermionic
supervielbein components related to Poincare supersymmetry obey
the transformation rules
\beq \delta_f\,\hat\omega^\mu_{\hat
a}(d)=j^\mu_{\hat a}df
\eeq
with the current contribution
\beq
\label{dilspoincarecf} j^{\mu}_{\hat
a}=\frac{\partial\hat\omega^\mu_{\hat a}(\delta_f)}{\partial
f}=e^{-\varphi}(\hat\Theta^\mu_{\hat a}-2\tilde
x^{\mu\nu}\hat\eta_{\nu\hat a}),
\eeq
while those related to
conformal supersymmetry transform as
\beq
\delta_f\hat\chi_{\mu\hat a}(d)=J_{\mu\hat a}df,
\eeq
where the
corresponding current contribution is given by
\beq
\label{dilsconfcf} J_{\mu\hat a}=\frac{\partial\hat\chi_{\mu\hat
a}(\delta_f)}{\partial
f}=-e^{\varphi}\left(\Lambda_{-\mu}{}^\nu\hat\eta_{\nu\hat
a}+ie^{\varphi}(\bar\eta\eta)\varepsilon_{\mu\nu}j^\nu_{\hat a}\right).
\eeq

Above presented current contributions determine the Noether current density related to the scale invariance of superstring action (\ref{action})
\beq
\mathcal J^{i}(\tau,\sigma)=\mathcal J_{AdS}{}^i+\mathcal J_{CP}{}^{i}+\mathcal J_{WZ}{}^{i}.
\eeq
Specifically the $AdS$ part of the current density
\beq
\mathcal J_{AdS}{}^{i}=-\sqrt{-g}g^{ij}\left(\frac{1}{4}(\hat\omega_{jm}+\hat
c_{jm})j^{m}+\Delta_jj\right)
\eeq
is contributed by Eq.(\ref{diladscf}), the $\mathbb{CP}^3$ part
\beq
\mathcal J_{CP}{}^{i}=\frac12\sqrt{-g}g^{ij}\left(\Omega_{ja}{}^4\;{}^*\!j^a+\Omega_{j4}{}^a\;{}^*\!\bar j_a\right)
\eeq
is determined by the current contributions that enter (\ref{dilcp3viel}), and the WZ part
\beq
\mathcal J_{WZ}{}^{i}=\frac{i}{4}\varepsilon^{ij}\mathfrak J_{\hat a}{}^{\hat b}\left(\hat{\omega}^{\mu\hat a}_j\varepsilon_{\mu\nu}j^{\nu}_{\hat b}+\hat{\chi}^{\hat a}_{j\mu}\varepsilon^{\mu\nu}J_{\nu\hat b}\right)
\eeq
receives contributions from (\ref{dilspoincarecf}) and (\ref{dilsconfcf}).

\subsubsection{Lorentz rotations}

Under local $SO(1,2)$ Lorentz rotations with parameters $l^{mn}$ Cartan forms identified with the supervielbein components tangent to the $AdS_4$ space exhibit the following transformation properties
\beq
\begin{array}{rl}
\delta_{l}\hat\omega^m(d)+\delta_{l}\hat c^m(d) &=j^{m}{}_{kn}dl^{kn}+l^{mn}(\hat\omega_{n}(d)+\hat c_{n}(d)),\\
\delta_l\Delta(d)&=j_{mn}dl^{mn}
\end{array}
\eeq
with the current contributions
\beq \label{lorentzadscf}
\begin{array}{rl}
j^{m}{}_{kn}=&\frac{\partial(\hat\omega^m(\delta_l)+\hat c^m(\delta_l))}{\partial l^{kn}}=\frac12e^{-2\varphi}A\left(\delta^m_kx_n-\delta^m_nx_k+i(\bar\theta\theta)\varepsilon^m{}_{kn}\right)\\[0.2cm]
+&\frac12e^{2\varphi}(\bar\eta\eta)\left\{\delta^m_k\left[(\eta_a\sigma_n\bar\theta^a)+(\bar\eta^a\sigma_n\theta_a)\right]-(k\leftrightarrow n)+\varepsilon^m{}_{kn}\left[i+(\bar\eta\theta)+(\bar\theta\eta)\right]\right\}
,\\[0.2cm]
j_{mn}=&\frac{\partial\Delta(\delta_l)}{\partial l^{mn}}=\frac{i}{4}\left[(\theta_a\sigma_{mn}\bar\eta^a)+(\bar\theta^a\sigma_{mn}\eta_a)\right].
\end{array}
\eeq
The $su(4)$ Cartan forms are obviously $D=3$ Lorentz invariant
\beq
\delta_l\Omega_{\hat a}{}^{\hat b}(d)=J_{\hat a}{}^{\hat b}{}_{mn}dl^{mn}
\eeq
modulo the current contribution matrix
\beq\label{su4ccmlorentz}
J_{\hat a}{}^{\hat b}{}_{mn}=\textstyle{\frac{\partial}{\partial l^{mn}}}\Omega_{\hat a}{}^{\hat b}(\delta_l)=\left(
\begin{array}{cc}
j_a{}^{b}{}_{mn} & j_{abmn}\\[0.2cm]
-\bar j^{ab}{}_{mn} & -j_b{}^{a}{}_{mn}
\end{array}\right)=\frac12\left[(\hat\Theta_{\hat a}\sigma_{mn}\hat\eta^{\hat b})-(\hat\Theta^{\hat b}\sigma_{mn}\hat\eta_{\hat a})\right].
\eeq
Hence variation of the supervielbein components tangent to the $\mathbb{CP}^3$ manifold is extracted from (\ref{su4ccmlorentz})
\beq \label{lorentzcp3viel}
\delta_l\Omega_{a}{}^4(d)=-{}^*\!\bar j_{amn}dl^{mn},\quad\delta_l\Omega_{4}{}^a(d)=-{}^*\! j^a{}_{mn}dl^{mn}.
\eeq
Variation of the supervielbein fermionic components that are identified with the Cartan forms related to Poincare supersymmetry follows from the general expression (\ref{genvarspoincarecf})
\beq
\delta_l\hat\omega^\mu_{\hat a}(d)=j^{\mu}_{\hat amn}dl^{mn}+\frac14\hat\omega^\nu_{\hat a}(d)l_{\nu}{}^\mu,
\eeq
where the current contribution reads
\beq \label{spoincarelorentz}
j^{\mu}_{\hat amn}=\frac{\partial\hat\omega^\mu_{\hat a}(\delta_l)}{\partial l^{mn}}=\frac14e^{-\varphi}\left(\hat\Theta^\nu_{\hat a}\sigma_{mn\nu}{}^\mu-\hat{\tilde x}{}^{\mu\lambda}\sigma_{mn\lambda}{}^\nu\hat\eta_{\nu\hat a}\right).
\eeq
Transformation properties of the Cartan forms related to conformal supersymmetry identified with another half supervielbein fermionic components follow from Eq.(\ref{genvarsconfcf})
\beq
\delta_l\hat\chi_{\mu\hat a}(d)=J_{\mu\hat amn}dl^{mn}-
\frac14l_{\mu}{}^\nu\hat\chi_{\nu\hat a}(d)
\eeq
with the current contribution
\beq \label{sconflorentz}
J_{\mu\hat amn}=\frac{\partial\hat\chi_{\mu\hat a}(\delta_l)}{\partial l^{mn}}=-e^{\varphi}\left({\textstyle\frac14}\Lambda_{-\mu}{}^\nu\sigma_{mn\nu}{}^\lambda\hat\eta_{\lambda\hat a}+ie^{\varphi}(\bar\eta\eta)\varepsilon_{\mu\nu}j^{\nu}_{\hat amn}\right).
\eeq

Then one is able to derive the current density related to global $SO(1,2)$ symmetry of the superstring action (\ref{action})
\beq
\mathcal J^{i}{}_{mn}(\tau,\sigma)=\mathcal J_{AdS}{}^{i}{}_{mn}+\mathcal J_{CP}{}^{i}{}_{mn}+\mathcal J_{WZ}{}^{i}{}_{mn}.
\eeq
The form of the $AdS$ part of the current density
\beq
\mathcal J_{AdS}{}^{i}{}_{mn}=-\sqrt{-g}g^{ij}\left(\frac{1}{4}(\hat\omega_{j\, l}+\hat
c_{j\, l})j^{l}{}_{mn}+\Delta_jj_{mn}\right),
\eeq
is determined by Eq.(\ref{lorentzadscf}), that of the $\mathbb{CP}^3$ part
\beq
\mathcal J_{CP}{}^{i}{}_{mn}=\frac12\sqrt{-g}g^{ij}\left(\Omega_{ja}{}^4\;{}^*\! j^a{}_{mn}+\Omega_{j4}{}^a\;{}^*\!\bar j_{amn}\right)
\eeq
by Eq.(\ref{lorentzcp3viel}),
and the WZ part of the current density
\beq
\mathcal J_{WZ}{}^{i}{}_{mn}=\frac{i}{4}\varepsilon^{ij}\mathfrak J_{\hat a}{}^{\hat b}\left(\hat{\omega}^{\mu\hat a}_j\varepsilon_{\mu\nu}j^{\nu}_{\hat bmn}+\hat{\chi}{}^{\hat a}_{j\mu}\varepsilon^{\mu\nu}J_{\nu\hat bmn}\right)
\eeq
is contributed by Eqs.(\ref{spoincarelorentz})
and (\ref{sconflorentz}).

\subsection{Noether currents associated with $SU(4)$ $R-$symmetry}

\subsubsection{$U(3)$ Rotations}

Global $U(3)$ rotations represent an obvious symmetry of the $AdS_4$ part of $(10|24)-$supervielbein thus the nontrivial part of its variation under the coordinate-dependent $U(3)$ rotations
\beq
\delta_w\hat\omega^m(d)+\delta_w\hat c^m(d)=j^m{}_a{}^bdw_b{}^a,\quad\delta_w\Delta=j_a{}^bdw_b{}^a
\eeq
is concentrated in the current contributions
\beq \label{u3adscf}
\begin{array}{rl}
j^{m}{}_a{}^b=&\frac{\partial(\hat\omega^m(\delta_w)+\hat c^m(\delta_w))}{\partial w_b{}^a}=2e^{-2\varphi}A\left[\delta_a^b(\theta_c\sigma^m\bar\theta^c)-(\theta_a\sigma^m\bar\theta^b)\right]
+2e^{2\varphi}\delta_a^b\left\{(\eta_c\tilde\sigma^m\bar\eta^c)\right.\\[0.2cm]
-&\left.i(\bar\eta\eta)\left[(\theta_c\sigma^m\bar\eta^c)+(\eta_c\sigma^m\bar\theta^c)\right]\right\}-2e^{2\varphi}\left\{(\eta_a\tilde\sigma^m\bar\eta^b)-i(\bar\eta\eta)\left[(\theta_a\sigma^m\bar\eta^b)+(\eta_a\sigma^m\bar\theta^b)\right]\right\},\\[0.2cm]
j_a{}^b=&\frac{\partial\Delta(\delta_w)}{\partial w_b{}^a}=\delta_a^b(\bar\theta^{\mu c}\eta_{\mu c}+\bar\eta^c_\mu\theta^\mu_c)+\theta^\mu_a\bar\eta^b_\mu+\eta_{\mu a}\bar\theta^{\mu b}.
\end{array}
\eeq
Transformation properties of the $su(4)$ Cartan forms under $U(3)$ rotations are derived from Eq.(\ref{genvarsu4cf})
\beq \label{wvarOmega}
\delta_w\Omega_{\hat c}{}^{\hat d}(d)=\hat J_{\hat c}{}^{\hat d}{}_a{}^bdw_b{}^a+i(\Omega_{\hat c}{}^{\hat e}(d)(\widehat W|_w)_{\hat e}{}^{\hat d}-(\widehat W|_w)_{\hat c}{}^{\hat e}\Omega_{\hat e}{}^{\hat d}(d))-d(\widehat W|_w)_{\hat c}{}^{\hat d}.
\eeq
The current contribution matrix
\beq
\hat J_{\hat c}{}^{\hat d}{}_a{}^b=\frac{\partial}{\partial w_b{}^a}\Omega_{\hat c}{}^{\hat d}(\delta_w)=\left(
\begin{array}{cc}
\hat j_c{}^d\vphantom{\hat j}_a\vphantom{\hat j}^b & \hat j_{cd}\vphantom{\hat j}_a\vphantom{\hat j}^b\\[0.2cm]
-\hat{\bar j}\vphantom{\hat{\bar j}}^{cd}\vphantom{\hat{\bar
j}}_a\vphantom{\hat j}^b & -\hat j_d{}^c\vphantom{\hat
j}_a\vphantom{\hat j}^b
\end{array}\right)
\eeq
is obtained by $T-$transformation of the matrix
\beq
\begin{array}{rl}
J_{\hat c}{}^{\hat d}{}_a{}^b=&\textstyle{\frac{\partial}{\partial w_b{}^a}}((\widetilde W|_w)_{\hat c}{}^{\hat d}+\Psi_{\hat c}{}^{\hat d}(\delta_w))=\delta_{\hat c}{}^b\delta_a{}^{\hat d}-\delta_{\hat ca}\delta^{b\hat d}+4\eta_{\mu\hat c}(\theta^\mu_a\bar\theta^{\nu b}+\theta^\nu_a\bar\theta^{\mu b})\eta^{\hat d}_\nu\\[0.2cm]
+&\delta_a^b\left[\frac{i}{2}\mathfrak J_{\hat c}{}^{\hat d}+(\mathfrak J\theta^\mu)_{\hat c}\eta^{\hat d}_\mu-(\mathfrak J\theta^\mu)^{\hat d}\eta_{\mu\hat c}-4\eta_{\mu\hat c}(\theta^\mu_e\bar\theta^{\nu e}+\theta^\nu_e\bar\theta^{\mu e})\eta^{\hat d}_\nu\right]\\[0.2cm]
+&2i\left[\theta^\mu_a\left(\eta_{\mu\hat c}\delta^{b\hat d}-\eta^{\hat d}_\mu\delta_{\hat c}{}^b\right)+\bar\theta^{\mu b}\left(\eta^{\hat d}_\mu\delta_{\hat ca}-\eta_{\mu\hat c}\delta_a{}^{\hat d}\right)\right],
\end{array}
\eeq
where the following objects have been introduced
\beq
\delta_{\hat c}{}^b=\left(
\begin{array}{c}
\delta^b_c\\
0
\end{array}\right),\quad\delta_{\hat ca}=\left(
\begin{array}{c}
0\\
\delta^c_a
\end{array}\right),\quad
\delta_a{}^{\hat d}=\left(\delta^d_a\ 0\right),\quad\delta^{b\hat d}=\left( 0\ \delta^b_d\right).
\eeq
So that the explicit form of the current contribution matrix is
\beq\label{u3su4ccm}
\hat J_{\hat c}{}^{\hat d}{}_a{}^b=(TJ_a{}^b\bar T)_{\hat c}{}^{\hat d}=\mathcal T_{\hat c}{}^b\mathcal T^{\hat d}{}_a-\mathcal T_{\hat ca}\mathcal T^{\hat db}+\frac{i}{2}\delta_a^b(\mathcal T\mathfrak J\bar{\mathcal T})_{\hat c}{}^{\hat d}.
\eeq
The matrix $\mathcal T_{\hat a}{}^{\hat b}$ equal
\beq
\mathcal T_{\hat a}{}^{\hat b}=\left(
\begin{array}{cc}
\mathcal T_a{}^b & \mathcal T_{ab}\\
\mathcal T^{ab} & \mathcal T^a{}_b
\end{array}
\right)=T_{\hat a}{}^{\hat b}+2i\hat\eta_{\nu\hat a}\theta^{\nu\hat b}
\eeq
will also be used below.
From Eqs.(\ref{wvarOmega}) and (\ref{u3su4ccm}) one derives transformation properties of the $\mathbb{CP}^3$ part of the bosonic supervielbein
\beq \label{u3cp3vielb}
\delta_w\Omega_c{}^4(d)=-({}^*\!\hat{\bar j}_{c})\vphantom{\hat{\bar j}}^{\vphantom{b}}_{a}\vphantom{\hat{\bar j}}^bdw_b{}^a+i(\hat w|_w)_d{}^d\Omega_c{}^4(d)-i(\hat w|_w)_c{}^d\Omega_d{}^4(d)
\eeq 
and c.c. expression. 

Fermionic supervielbein components associated with the Poincare
supersymmetry transform under local $U(3)$ rotations as follows
\beq
\delta_w\hat\omega^\mu_{\hat c}(d)=j^{\mu}_{\hat
c}\vphantom{j}^{\vphantom{\mu}}_{\vphantom{\hat c}a}\vphantom{j}^{b} dw_b{}^a-i(\widehat W|_w)_{\hat c}{}^{\hat
d}\,\hat\omega_{\hat d}^\mu(d),
\eeq
where
\beq
\label{u3spoincare} j^{\mu}_{\hat
c}\vphantom{j}^{\vphantom{\mu}}_{\vphantom{\hat c}a}\vphantom{j}^b=\frac{\partial\hat\omega^\mu_{\hat
c}(\delta_w)}{\partial
w_b{}^a}=e^{-\varphi}\left[{\textstyle\frac12}\delta_a^b(\mathcal
T\mathfrak J\theta^\mu)_{\hat c}-i\theta^\mu_a\mathcal T_{\hat
c}{}^b+i\bar\theta^{\mu b}\mathcal T_{\hat ca}\right]
\eeq
represents the current contribution. Analogously variation of the
fermionic supervielbein components related to conformal
supersymmetry is given by the expression
\beq
\delta_w\hat\chi_{\mu\hat c}(d)=J_{\mu\hat
c}{}_a{}^bdw_b{}^a-i(\widehat W|_w)_{\hat c}{}^{\hat
d}\hat\chi_{\mu\hat d}(d)
\eeq
with the current contribution
\beq
\label{u3sconf} J_{\mu\hat
c}{}_a{}^b=\frac{\partial\hat\chi_{\mu\hat c}(\delta_w)}{\partial
w_b{}^a}=e^{\varphi}\left[{\textstyle\frac12}\delta_a^b(\mathcal
T\mathfrak J\eta_\mu)_{\hat c}-i\eta_{\mu a}\mathcal T_{\hat
c}{}^b+i\bar\eta^b_\mu\mathcal T_{\hat
ca}+ie^{\varphi}(\bar\eta\eta)\varepsilon_{\mu\nu}j^\nu_{\hat c}\vphantom{j}^{\vphantom{\nu}}_a\vphantom{j}^b\right].
\eeq

In summary the current density associated with the $U(3)$ global
invariance of superstring action (\ref{action})
\beq
\mathcal
J^{i}{}_a{}^b(\tau,\sigma)=\mathcal
J_{AdS}{}^{i}{}_{a}{}^b+\mathcal J_{CP}{}^{i}{}_{a}{}^b+\mathcal
J_{WZ}{}^{i}{}_{a}{}^b
\eeq
consists of three summands
\beq
\mathcal
J_{AdS}{}^{i}{}_{a}{}^b=-\sqrt{-g}g^{ij}\left(\frac{1}{4}(\hat\omega_{jm}+\hat
c_{jm})j^{m}{}_a{}^b+\Delta_jj_{a}{}^b\right),
\eeq
\beq
\mathcal
J_{CP}{}^{i}{}_{a}{}^b=\frac12\sqrt{-g}g^{ij}\left(\Omega_{jc}{}^4\;({}^*\!\hat
j^c)\vphantom{\hat j}_a\vphantom{\hat
j}^b+\Omega_{j4}{}^c\;({}^*\!\hat{\bar j}_{c})\vphantom{\hat{\bar
j}}^{\vphantom{b}}_a\vphantom{\hat{\bar j}}^b\right),
\eeq
and
\beq
\mathcal
J_{WZ}{}^{i}{}_{a}{}^b=\frac{i}{4}\varepsilon^{ij}\mathfrak
J_{\hat c}{}^{\hat d}\left(\hat{\omega}^{\mu\hat
c}_j\varepsilon_{\mu\nu}j^{\nu}_{\hat d}\vphantom{j}^{\vphantom{\nu}}_{\vphantom{\hat d}a}\vphantom{j}^b+\hat{\chi}^{\hat
c}_{j\mu}\varepsilon^{\mu\nu}J_{\nu\hat d}\vphantom{J}_{\vphantom{\hat d}a}{}^b\right)
\eeq
that are determined by the current contributions of the
$osp(4|6)/(so(1,3)\times u(3))$ Cartan forms (\ref{u3adscf}),
(\ref{u3cp3vielb}), (\ref{u3spoincare}) and (\ref{u3sconf}).

\subsubsection{$SU(4)/U(3)$ transformations}

As in the case of $U(3)$ rotations Cartan forms from the $AdS_4$ sector are invariant under local $SU(4)/U(3)$ transformations
\beq
\delta_y\hat\omega^m(d)+\delta_y\hat c^m(d)=j^m{}_ady^a+\bar j^{ma}d\bar y_a,\quad\delta_y\Delta(d)=j_ady^a+\bar j^ad\bar y_a
\eeq
modulo the current contributions
\beq \label{su4overu3adscf}
\begin{array}{rl}
j^m{}_a=&\frac{\partial(\hat\omega^m(\delta_y)+\hat c^m(\delta_y))}{\partial y^a}=-\varepsilon_{abc}\left\{e^{-2\varphi}A(\bar\theta^b\sigma^m\bar\theta^c)+e^{2\varphi}\left[(\bar\eta^b\tilde\sigma^m\bar\eta^c)-2i(\bar\eta\eta)(\bar\theta^b\tilde\sigma^m\bar\eta^c)\right]\right\},\\[0.2cm]
j_a=&\frac{\partial\Delta(\delta_y)}{\partial y^a}=\varepsilon_{abc}\bar\theta^{\mu b}\bar\eta^c_\mu 
\end{array}
\eeq 
and c.c. 
Transformation properties of the
$su(4)$ Cartan forms follow from the general formula
(\ref{genvarsu4cf}) by specializing to $SU(4)/U(3)$ rotations
\beq
\label{yvarOmega}
\delta_y\Omega_{\hat b}{}^{\hat c}(d)=J_{\hat
b}\vphantom{J}^{\hat c}\vphantom{J}_{\vphantom{\hat b}a}dy^a+\bar J_{\hat b}{}^{\hat ca}d\bar
y_a+i\left(\Omega_{\hat b}{}^{\hat d}(d)(\widehat W|_y)_{\hat
d}{}^{\hat c}-(\widehat W|_y)_{\hat b}{}^{\hat d}\Omega_{\hat
d}{}^{\hat c}(d)\right)-d(\widehat W|_y)_{\hat b}{}^{\hat c}.
\eeq
Corresponding current contribution matrices equal
\beq
J_{\hat
b}\vphantom{J}^{\hat c}\vphantom{J}_{\vphantom{\hat b}a}=\textstyle{\frac{\partial}{\partial
y^a}}\Omega_{\hat b}{}^{\hat c}(\delta_y)=\left(
\begin{array}{cc}
j_b{}^{c}{}_a & j_{bca}\\[0.2cm]
j^{bc}{}_a & -j_c{}^{b}{}_a
\end{array}\right)=-\varepsilon_{ade}\mathcal T_{\hat b}{}^d\mathcal T^{\hat ce}
\eeq
and
\beq
\bar J_{\hat b}{}^{\hat c}{}^a=\textstyle{\frac{\partial}{\partial\bar y_a}}\Omega_{\hat b}{}^{\hat c}(\delta_y)=
\left(
\begin{array}{cc}
\bar j_b{}^{c}{}^a & -\bar j_{bc}{}^{a}\\[0.2cm]
-\bar j^{bca} & -\bar j_c{}^{b}{}^a
\end{array}\right)=\varepsilon^{ade}\mathcal T_{\hat bd}\mathcal T^{\hat c}{}_{e}.
\eeq
So that one extracts from the above expressions the transformation rules for the supervielbein bosonic components tangent to the $\mathbb{CP}^3$ manifold
\beq \label{ycp3vielb}
\delta_y\Omega_b{}^4(d)=({}^*\! j_b){}_ady^a-({}^*\!\bar j_b){}^ad\bar y_a+i(\hat w|_y)_c{}^c\Omega_b{}^4(d)-i(\hat w|_y)_b{}^c\Omega_c{}^4(d).
\eeq
Supervielbein fermionic components associated with the Poincare supersymmetry have the following properties under $SU(4)/U(3)$ transformations
\beq
\delta_y\hat\omega^\mu_{\hat b}(d)=j^{\mu}_{\hat ba}dy^a+\bar j^{\mu a}_{\hat b}d\bar y_a-i(\widehat W|_y)_{\hat b}{}^{\hat c}\,\hat\omega_{\hat c}^\mu(d)
\eeq
with the current contributions
\beq \label{spoincaresu4overu3}
j^{\mu}_{\hat ba}=\frac{\partial\hat\omega^\mu_{\hat b}(\delta_y)}{\partial y^a}=ie^{-\varphi}\varepsilon_{acd}\mathcal T_{\hat b}{}^c\bar\theta^{\mu d}
\eeq 
and c.c. 
Supervielbein fermionic components related to conformal supersymmetry transform as
\beq
\delta_y\hat\chi_{\mu\hat b}(d)=J_{\mu\hat b}\vphantom{J}_{\vphantom{\hat b}a}dy^a+\bar J_{\mu\hat b}{}^{a}d\bar y_a-i(\widehat W|_y)_{\hat b}{}^{\hat c}\hat\chi_{\mu\hat c}(d),
\eeq
where the current contributions can be brought to the form
\beq \label{sconfsu4overu3}
J_{\mu\hat b}\vphantom{J}_{\vphantom{\hat b}a}=\frac{\partial\hat\chi_{\mu\hat b}(\delta_y)}{\partial y^a}=ie^{\varphi}(\varepsilon_{acd}\mathcal T_{\hat b}{}^{c}\bar\eta_{\mu}^{d}+e^{\varphi}(\bar\eta\eta)\varepsilon_{\mu\nu}j^\nu_{\hat ba})
\eeq 
and c.c. 

Above derived current contributions of the supervielbein components determine the Noether current density associated with $SU(4)/U(3)$ global invariance of the superstring action
\beq
\mathcal J^{i}{}_a(\tau,\sigma)=\mathcal J_{AdS}{}^{i}{}_{a}+\mathcal J_{CP}{}^{i}{}_{a}+\mathcal J_{WZ}{}^{i}{}_{a} 
\eeq
and c.c. expression. The summands contributed by Eqs.(\ref{su4overu3adscf}), (\ref{ycp3vielb}), (\ref{spoincaresu4overu3}) and (\ref{sconfsu4overu3}) respectively take the form
\beq
\mathcal J_{AdS}{}^{i}{}_{a}=-\sqrt{-g}g^{ij}\left(\frac{1}{4}(\hat\omega_{jm}+\hat
c_{jm})j^{m}{}_a+\Delta_jj_{a}\right),
\eeq
\beq
\mathcal J_{CP}{}^{i}{}_{a}=\frac12\sqrt{-g}g^{ij}\left(\Omega_{jb}{}^4\;({}^*\! j^b){}_a-\Omega_{j4}{}^b\;({}^*\! j_b){}_a\right),
\eeq
and
\beq
\mathcal J_{WZ}{}^{i}{}_{a}=\frac{i}{4}\varepsilon^{ij}\mathfrak J_{\hat b}{}^{\hat c}\left(\hat{\omega}^{\mu\hat b}_j\varepsilon_{\mu\nu}j^{\nu}_{\hat ca}+\hat{\chi}^{\hat b}_{j\mu}\varepsilon^{\mu\nu}J_{\nu\hat c}{}_a\right).
\eeq

\subsection{Noether currents associated with $D=3$ $\mathcal N=6$ Poincare supersymmetry}

The superstring action (\ref{action}) is manifestly invariant under Poincare supersymmetry as $D=3$ $\mathcal N=6$ supercoordinates $(x^m, \theta^\mu_a, \bar\theta^{\mu a})$ that are the only non-trivially transforming ones enter through the supersymmetric Volkov-Akulov 1-forms \cite{VA72}. Hence non-invariance of the $AdS$ part of the supervielbein bosonic components under coordinate-dependent Poincare supersymmetry transformations
\beq
\delta_\varepsilon\hat\omega^m(d)+\delta_\varepsilon\hat c^m(d)=j^{m}{}_\mu^ad\varepsilon^{\mu}_{a}-\bar j^{m}{}_{\mu a}d\bar\varepsilon^{\mu a},\quad\delta_\varepsilon\Delta(d)=j^{a}_\mu d\varepsilon^{\mu}_{a}-\bar j_{\mu a}d\bar\varepsilon^{\mu a},
\eeq
is accounted by the current contributions which explicit form is
\beq \label{susyadscf}
\begin{array}{rl}
j^{m}{}_\mu^a=&\frac{\partial(\hat\omega^m(\delta_\varepsilon)+\hat c^m(\delta_\varepsilon))}{\partial\varepsilon^{\mu}_{a}}
=2\sigma^m_{\mu\nu}\left[ie^{-2\varphi}A\bar\theta^{\nu a}+e^{2\varphi}(\bar\eta\eta)\bar\eta^{\nu a}\right],\\[0.2cm]
j^{a}_\mu=&\frac{\partial\Delta(\delta_\varepsilon)}{\partial\varepsilon^{\mu}_{a}}=-i\bar\eta^a_\mu 
\end{array}
\eeq 
and c.c. expressions. 
The $su(4)$ Cartan forms are also $D=3$ $\mathcal N=6$
super-Poincare invariant
\beq
\delta_\varepsilon\Omega_{\hat
b}{}^{\hat c}(d)=J_{\hat b}\vphantom{J}^{\hat
c}\vphantom{J}^{a}_\mu d\varepsilon^{\mu}_{a} -\bar J_{\hat
b}\vphantom{\bar J}^{\hat c}\vphantom{\bar J}_{\mu
a}d\bar\varepsilon^{\mu a}
\eeq
modulo the current contribution
matrices
\beq
J_{\hat b}\vphantom{J}^{\hat
c}\vphantom{J}_{\mu}^{a}=\textstyle{\frac{\partial}{\partial\varepsilon^{\mu}_{a}}}\Omega_{\hat
b}{}^{\hat c}(\delta_\varepsilon)=\left(
\begin{array}{cc}
j_b\vphantom{j}^c\vphantom{j}_{\mu}^{a} & j^{\vphantom{a}}_{bc}\vphantom{j}_{\mu}^{a}\\
j^{bc}\vphantom{j}_{\mu}^{a} &
-j^{\vphantom{b}}_c\vphantom{j}^b_{\vphantom{c}}\vphantom{j}_{\mu}^{a}
\end{array}\right)
=2(\hat\eta_{\mu\hat b}\mathcal T^{\hat ca}-\mathcal T_{\hat
b}{}^a\hat\eta_{\mu}^{\hat c})
\eeq
and
\beq \bar J_{\hat
b}\vphantom{J}^{\hat c}\vphantom{J}_{\mu
a}=-\textstyle{\frac{\partial}{\vphantom{\hat\xi}\partial\bar\varepsilon^{\mu
a}}}\Omega_{\hat b}{}^{\hat c}(\delta_\varepsilon)=\left(
\begin{array}{cc}
\bar j_b\vphantom{\bar j}^c\vphantom{\bar j}_{\mu a} & -\bar j_{bc\mu a}\\
-\bar j^{bc}\vphantom{\bar j}_{\mu a} & -\bar j_c\vphantom{\bar
j}^b\vphantom{\bar j}_{\mu a}
\end{array}\right)
=2(\mathcal T_{\hat ba}\hat\eta{}_\mu^{\hat c}-\hat\eta_{\mu\hat b}\mathcal T^{\hat c}{}_{a})
\eeq
so that the $\mathbb{CP}^3$ part of the bosonic supervielbein transforms as
\beq \label{susycp3cf}
\delta_\varepsilon\Omega_b{}^4(d)=({}^*\! j_{b}){}_{\mu}^ad\varepsilon^\mu_a+({}^*\!\bar j_b){}_{\mu a}d\bar\varepsilon^{\mu a},
\quad\delta_\varepsilon\Omega_4{}^b(d)=-({}^*\! j^b){}_\mu^ad\varepsilon^\mu_a-({}^*\!\bar j^b){}_{\mu a}d\bar\varepsilon^{\mu a}.
\eeq
Cartan forms associated with the Poincare supersymmetry are manifestly $D=3$ $\mathcal N=6$ super-Poincare invariant
\beq
\delta_\varepsilon\hat\omega^\nu_{\hat b}(d)=j^\nu_{\hat b}\vphantom{j}^a_\mu d\varepsilon^\mu_a+\bar j^\nu_{\hat b}\vphantom{\bar j}^{\vphantom{\nu}}_{\mu a}d\bar\varepsilon^{\mu a}
\eeq
up to the current contributions
\beq \label{susyspoincarecf}
j^\nu_{\hat b}\vphantom{j}^a_\mu=\textstyle{\frac{\partial\hat\omega^\nu_{\hat b}(\delta_\varepsilon)}{\partial\varepsilon^{\mu}_{a}}}=e^{-\varphi}\left(\delta_\mu^\nu\mathcal T_{\hat b}{}^a+2i\hat\eta_{\mu\hat  b}\bar\theta^{\nu a}\right)
\eeq 
and c.c., 
as well as Cartan forms associated with the conformal supersymmetry
\beq
\delta_\varepsilon\hat\chi_{\nu\hat b}(d)=J_{\nu\hat b}\vphantom{J}^a_\mu d\varepsilon^\mu_a+\bar J_{\nu\hat b\mu a}d\bar\varepsilon^{\mu a}
\eeq
with the corresponding current contributions given by
\beq \label{susysconfcf}
J_{\nu\hat b}\vphantom{J}^a_\mu=\textstyle{\frac{\partial\hat\chi_{\nu\hat b}(\delta_\varepsilon)}{\partial\varepsilon^{\mu}_{a}}}=ie^{\varphi}\left(2\hat\eta_{\mu\hat b}\bar\eta^a_\nu-e^{\varphi}(\bar\eta\eta)\varepsilon_{\nu\lambda}j^\lambda_{\hat b}{}^a_\mu\right) 
\eeq
and c.c. 

Putting all together contributions (\ref{susyadscf}), (\ref{susycp3cf}), (\ref{susyspoincarecf}) and
(\ref{susysconfcf}) allows to determine the current density related to $D=3$ $\mathcal N=6$ supersymmetry invariance of the superstring action (\ref{action})
\beq\label{spoincarecurdens}
\mathcal J^{i}{}^a_\mu(\tau,\sigma)=\mathcal J_{AdS}{}^{i}{}^a_\mu+\mathcal J_{CP}{}^{i}{}^a_\mu+\mathcal J_{WZ}{}^{i}{}^a_\mu
\eeq
and the c.c. one.  
The individual summands entering the current density (\ref{spoincarecurdens}) equal
\beq
\mathcal J_{AdS}{}^{i}{}^a_{\mu}=-\sqrt{-g}g^{ij}\left(\frac{1}{4}(\hat\omega_{jm}+\hat
c_{jm})j^{m}{}^a_{\mu}+\Delta_jj^a_{\mu}\right),
\eeq
\beq
\mathcal J_{CP}{}^{i}{}^a_{\mu}=\frac12\sqrt{-g}g^{ij}\left(\Omega_{jb}{}^4\;({}^*\! j^b){}^a_\mu-\Omega_{j4}{}^b\;({}^*\! j_b){}^a_{\mu}\right)
\eeq
and
\beq
\mathcal J_{WZ}{}^{i}{}^a_{\mu}=\frac{i}{4}\varepsilon^{ij}\mathfrak J_{\hat b}{}^{\hat c}\left(\hat{\omega}^{\nu\hat b}_j\varepsilon_{\nu\lambda}j^{\lambda}_{\hat c}\vphantom{j}^a_{\mu}+\hat{\chi}^{\hat b}_{j\nu}\varepsilon^{\nu\lambda}J^{\vphantom{a}}_{\lambda\hat c}\vphantom{J}^a_{\vphantom{\lambda\hat c}\mu}\right).
\eeq

\subsection{Noether currents associated with $D=3$ $\mathcal N=6$ conformal supersymmetry}

Transformation properties of the supervielbein bosonic components in the directions tangent to the $AdS_4$ space can be obtained from the general expressions (\ref{genvaradscf}) by appropriately restricting the parameters of $SO(1,3)$ compensating rotations
\beq
\begin{array}{rl}
\delta_\xi\hat\omega^m(d)+\delta_\xi\hat c^m(d)=&j^{m\mu a}d\xi_{\mu a}-\bar j^{m}{}^{\mu}_ad\bar\xi^a_\mu
+(\hat l|_\xi)^{mn}(\hat\omega_{n}(d)+\hat c_{n}(d))+4(\hat b|_\xi)^m\Delta(d),\\[0.2cm]
\delta_\xi\Delta(d)=&j^{\mu a}d\xi_{\mu a}-\bar j^{\mu}_ad\bar\xi^a_\mu-(\hat
b|_\xi)^m(\hat\omega_{m}(d)+\hat c_{m}(d))
\end{array}
\eeq
and adding the current contribution terms
\beq\label{sconfadsccm}
\begin{array}{rl}
j^{m\mu a}=&\frac{\partial(\hat\omega^m(\delta_\xi)+\hat c^m(\delta_\xi))}{\partial\xi_{\mu a}}
=2\left[ie^{-2\varphi}A\bar\theta^{\lambda
a}\sigma^m_{\lambda\nu}\hat{\tilde
x}{}^{\nu\mu}+ie^{2\varphi}\bar\eta^a_\lambda\tilde\sigma^{m\lambda\nu}\Lambda_{-\nu}{}^{\mu}\right.\\[0.2cm]
+&\left.e^{2\varphi}(\bar\eta\eta)\left(\bar\eta^{\lambda
a}\sigma^m_{\lambda\nu}\hat{\tilde
x}{}^{\nu\mu}-\bar\theta^a_\lambda\tilde\sigma^{m\lambda\nu}\Lambda_{+\nu}{}^\mu\right)\right],\\[0.2cm]
j^{\mu
a}=&\frac{\partial\Delta(\delta_\xi)}{\partial\xi_{\mu
a}}=-i\left(\bar\theta^{\nu
a}\Lambda_{-\nu}{}^\mu+\bar\eta^a_\nu\hat{\tilde
x}{}^{\nu\mu}\right) 
\end{array}
\eeq
and c.c. 

Variation of the matrix of $su(4)$ Cartan forms under the coordinate-dependent conformal supersymmetry transformations is presented in the following form
\beq \label{su4varsconf}
\delta_\xi\Omega_{\hat b}{}^{\hat c}(d)=\hat J_{\hat b}{}^{\hat c}{}^{\mu a}d\xi_{\mu a}
-\hat{\bar J}^{\vphantom{\mu}}_{\hat b}\vphantom{\hat{\bar J}}^{\hat c}_{\vphantom{a}}\vphantom{\hat{\bar J}}^\mu_ad\bar\xi_{\mu}^{a}+i\left(\Omega_{\hat b}{}^{\hat d}(d)(\widehat W|_\xi)_{\hat d}{}^{\hat c}-(\widehat W|_\xi)_{\hat b}{}^{\hat d}\Omega_{\hat d}{}^{\hat c}(d)\right)-d(\widehat W|_\xi)_{\hat b}{}^{\hat c},
\eeq
where the current contributions
\beq \label{sconfsu4ccm}
\hat J_{\hat b}{}^{\hat c}{}^{\mu a}=\textstyle{\frac{\partial}{\partial\xi_{\mu a}}}\Omega_{\hat b}{}^{\hat c}(\delta_\xi)=\left(
\begin{array}{cc}
\hat j_b{}^c{}^{\mu a} & \hat j_{bc}{}^{\mu a}\\
\hat j^{bc\mu a} & -\hat j_c{}^b{}^{\mu a}\end{array}\right)
\eeq
and
\beq \label{sconfsu4ccm'}
\hat{\bar J}^{\vphantom{\mu}}_{\hat b}\vphantom{\hat{\bar J}}^{\hat c}_{\vphantom{a}}\vphantom{\hat{\bar J}}^\mu_a=-\textstyle{\frac{\partial}{\vphantom{\hat\xi}\partial\bar\xi_{\mu}^{a}}}\Omega_{\hat b}{}^{\hat c}(\delta_\xi)=\left(
\begin{array}{cc}
\hat{\bar j}_b{}^c{}^{\mu}_{a} & -\hat{\bar j}_{bc}{}^{\mu}_{a}\\
-\hat{\bar j}{}^{bc}{}^{\mu}_{a} & -\hat{\bar j}_c{}^b{}^{\mu}_{a}\end{array}\right)
\eeq
are obtained by the $T-$transformation of the matrices
\beq
\begin{array}{c}
J_{\hat b}{}^{\hat c}{}^{\mu a}=\textstyle{\frac{\partial}{\partial\xi_{\mu a}}}((\widetilde W|_\xi)_{\hat b}{}^{\hat c}+\Psi_{\hat b}{}^{\hat c}(\delta_\xi))=2(\delta_{\hat b}{}^a\Theta^{\mu\hat c}-\delta^{a\hat c}\Theta^\mu_{\hat b})-4i\bar\theta^{\nu a}(\eta_{\nu\hat b}\Theta^{\mu\hat c}+\Theta^\mu_{\hat b}\eta^{\hat c}_\nu),\\[0.3cm]
\bar
J_{\hat b}{}^{\hat c}\vphantom{\bar J}^\mu_a=-\textstyle{\frac{\partial}{\vphantom{\hat\xi}\partial\bar\xi_{\mu}^{a}}}((\widetilde W|_\xi)_{\hat b}{}^{\hat c}+\Psi_{\hat b}{}^{\hat c}(\delta_\xi))=2(\delta_a{}^{\hat c}\Theta^\mu_{\hat b}-\delta_{\hat ba}\Theta^{\mu\hat c})+4i\theta^\nu_a(\eta_{\nu\hat b}\Theta^{\mu\hat c}+\Theta^\mu_{\hat b}\eta^{\hat c}_\nu).
\end{array}
\eeq
So that the current contribution matrices (\ref{sconfsu4ccm}) and (\ref{sconfsu4ccm'}) acquire the form
\beq
\hat J_{\hat b}{}^{\hat c}{}^{\mu a}
=(TJ^{\mu a}\bar T)_{\hat b}{}^{\hat c}=2(\mathcal T_{\hat b}{}^a\hat\Theta^{\mu\hat c}-\hat\Theta^{\mu}_{\hat b}\mathcal T^{\hat ca})
\eeq
and
\beq
\hat{\bar J}^{\vphantom{\mu}}_{\hat b}\vphantom{\hat{\bar J}}^{\hat c}_{\vphantom{a}}\vphantom{\hat{\bar J}}^\mu_a=(T\bar J^\mu_a\bar T)_{\hat b}{}^{\hat c}=2(\hat\Theta^\mu_{\hat b}\mathcal T^{\hat c}{}_{a}-\mathcal T_{\hat ba}\hat\Theta^{\mu\hat c}).
\eeq
Then the variation under conformal supersymmetry of the $\mathbb{CP}^3$ components of the supervielbein is brought to the form
\beq\label{sconfvarcp3viel}
\delta_\xi\Omega_{b}{}^4(d)=({}^*\!\hat j_b)\vphantom{\hat j}^{\mu a}d\xi_{\mu
a}+({}^*\!\hat{\bar j}_{b})\vphantom{\hat{\bar j}}^\mu_ad\bar\xi^a_\mu+i(\hat w|_\xi)_c{}^c\Omega_b{}^4(d)-i(\hat w|_\xi)_b{}^c\Omega_c{}^4(d) 
\eeq 
and c.c. expression. 

The variation of Cartan forms associated with the super-Poincare generators can be extracted from the general expression (\ref{genvarspoincarecf})
\beq
\delta_\xi\hat\omega_{\hat b}^\nu(d)=j^\nu_{\hat b}{}^{\mu a}d\xi_{\mu a}+\bar j^{\nu\mu}_{\hat ba}d\bar\xi^a_\mu+
\frac14\hat\omega^\lambda_{\hat b}(d)(\hat l|_\xi)_{\lambda}{}^\nu+(\hat{\tilde b}|_\xi)^{\nu\lambda}\hat\chi_{\lambda\hat b}(d)-i(\widehat W|_\xi)_{\hat b}{}^{\hat c}\hat\omega_{\hat c}^\nu(d)
\eeq
with the current contributions given by
\beq\label{sconfspoincarecc}
j^\nu_{\hat b}{}^{\mu a}=\textstyle{\frac{\partial\hat\omega^\nu_{\hat b}(\delta_\xi)}{\partial\xi_{\mu a}}}=e^{-\varphi}(\mathcal T_{\hat b}{}^a\hat{\tilde x}{}^{\nu\mu}+2i\bar\theta^{\nu a}\hat\Theta^\mu_{\hat b}) 
\eeq 
and c.c. 
Similarly the variation of Cartan forms associated with the conformal supersymmetry generators reads
\beq \label{sconfvarsconf}
\delta_\xi\hat\chi_{\nu\hat b}(d)=J_{\nu\hat b}{}^{\mu a}d\xi_{\mu a}+\bar J_{\nu\hat b}\vphantom{\bar J}^\mu_{a}d\bar\xi^a_\mu-
\frac14(\hat l|_\xi)_{\nu}{}^\lambda\hat\chi_{\lambda\hat b}(d)+(\hat b|_\xi)_{\nu\lambda}\hat\omega^\lambda_{\hat b}(d)-i(\widehat W|_\xi)_{\hat b}{}^{\hat c}\hat\chi_{\nu\hat c}(d)
\eeq
with the corresponding current contributions
\beq\label{sconfsconfcc}
J_{\nu\hat b}{}^{\mu a}=\textstyle{\frac{\partial\hat\chi_{\nu\hat b}(\delta_\xi)}{\partial\xi_{\mu a}}}=e^{\varphi}(\Lambda_{-\nu}{}^\mu\mathcal T_{\hat b}{}^a+2i\bar\eta^a_\nu\hat\Theta^\mu_{\hat b}-ie^{\varphi}(\bar\eta\eta)\varepsilon_{\nu\lambda}j^\lambda_{\hat b}{}^{\mu a})
\eeq 
and c.c. 

As a result current density associated with the superstring action (\ref{action}) invariance under conformal supersymmetry takes the form
\beq
\mathcal J^{i\mu a}(\tau,\sigma)=\mathcal J_{AdS}{}^{i\mu a}+\mathcal J_{CP}{}^{i\mu a}+\mathcal J_{WZ}{}^{i\mu a} 
\eeq 
and c.c. one. 
Current contributions (\ref{sconfadsccm}) enter the AdS part of the
current density
\beq\label{sconf-ads}
\mathcal J_{AdS}{}^{i\mu a}=
-\sqrt{-g}g^{ij}\left(\frac{1}{4}(\hat\omega_{jm}+\hat
c_{jm})j^{m\mu a}+\Delta_jj^{\mu a}\right),
\eeq
those entering Eq.(\ref{sconfvarcp3viel}) determine the $\mathbb{CP}^3$ part of the superconformal current density
\beq\label{sconf-cp}
\mathcal J_{CP}{}^{i\mu a}=\frac12\sqrt{-g}g^{ij}(\Omega_{jb}{}^4\;({}^*\!\hat j^{b}){}^{\mu a}-\Omega_{j4}{}^b\;({}^*\!\hat j_{b}){}^{\mu a}).
\eeq
The form of the WZ term contribution is obtained by substituting Eqs.(\ref{sconfspoincarecc}) and (\ref{sconfsconfcc})
\beq\label{sconf-wz}
\mathcal J_{WZ}{}^{i\mu a}=\frac{i}{4}\varepsilon^{ij}\mathfrak J_{\hat b}{}^{\hat c}(\hat\omega^{\nu\hat b}_j\varepsilon_{\nu\lambda}j^\lambda_{\hat c}{}^{\mu a}+\hat\chi^{\hat b}_{j\nu}\varepsilon^{\nu\lambda}J_{\lambda\hat c}{}^{\mu a}).
\eeq

\section{Conclusion}

The $OSp(4|6)/(SO(1,3)\times U(3))$ supercoset sigma-model \cite{AF08}, \cite{Stefanski} describes manifestly classically integrable part of the $AdS_4\times\mathbb{CP}^3$ superstring action \cite{GSWnew}. By virtue of the isomorphism between the $osp(4|6)$ superalgebra and $D=3$ $\mathcal N=6$ superconformal algebra it can be presented in the conformal basis for $osp(4|6)/(so(1,3)\times u(3))$ Cartan forms \cite{U08} that are identified with the 'reduced' $(10|24)-$dimensional $D=10$ $\mathcal N=2A$ superspace vielbein obtained from the full one \cite{GSWnew} by setting to zero 8 fermionic coordinates related to space-time supersymmetries broken by the $AdS_4\times\mathbb{CP}^3$ superbackground. The $OSp(4|6)/(SO(1,3)\times U(3))$ supercoset sigma-model action is by construction invariant under the global $OSp(4|6)$ supergroup transformations and hence is also invariant under $D=3$ $\mathcal N=6$ superconformal symmetry that is the global symmetry of ABJM gauge theory \cite{BLS}. In this paper we have derived explicit expressions for the corresponding world-sheet current densities associated with each type of the transformations from $D=3$ $\mathcal N=6$ superconformal symmetry. Considering the $OSp(4|6)/(SO(1,3)\times U(3))$ supercoset element parametrized by $D=3$ $\mathcal N=6$ super-Poincare coordinates supplemented by the $\mathbb{CP}^3$ coordinates, $AdS_4$ space bulk coordinate and Grassmann coordinates related to the conformal supersymmetry we have found their full transformations under $D=3$ $\mathcal N=6$ superconformal symmetry. So that passing to the canonical formulation it should be possible to calculate the algebra of associated supercharges.

Among the potential applications of the obtained results one could mention the semiclassical quantization around solutions to the $OSp(4|6)/(SO(1,3)\times U(3))$ superstring equations of motion \cite{semiclass}. They are also the starting point to examine residual symmetry algebras surviving upon fixing the gauge symmetries of $OSp(4|6)/(SO(1,3)\times U(3))$ sigma-model action (see, e.g.  \cite{Bykov}).

\section{Acknowledgements}

It is a pleasure to thank A.A.~Zheltukhin for interesting discussions and the Abdus Salam ICTP, where this work has been finalized, for warm hospitality and support.


\begin{thebibliography}{99}
\bibitem{Maldacena97}
J.M.~Maldacena, "The large N limit of superconformal field theories and supergravity", Adv.\ Theor.\ Math.\ Phys.\ \textbf{2} (1998) 231, \href{http://arxiv.org/abs/hep-th/9711200}{\texttt{arXiv:hep-th/9711200}}.

\bibitem{GKP98}
S.S.~Gubser, I.R.~Klebanov and A.M.~Polyakov, "Gauge theory correlators from non-critical string theory", Phys.\ Lett. \textbf{B428} (1998) 105, \href{http://arxiv.org/abs/hep-th/9802109}{\texttt{arXiv:hep-th/9802109}}.

\bibitem{Witten98}
E.~Witten, "Anti-de Sitter space and holography", Adv.\ Theor.\ Math.\ Phys.\ \textbf{2} (1998) 253, \href{http://arxiv.org/abs/hep-th/9802150}{\texttt{arXiv:hep-th/9802150}}.

\bibitem{ABJM}
O.~Aharony, O.~Bergman, D.L.~Jafferis and J.~Maldacena, "$\mathcal
N=6$ superconformal Chern-Simons-matter theories, M2-branes and
their gravity duals", JHEP \textbf{0810} (2008) 091,
\href{http://arxiv.org/abs/0806.1218}{\texttt{arXiv:0806.1218
[hep-th]}}.

\bibitem{MT98}
R.R.~Metsaev and A.A.~Tseytlin, "Type IIB superstring action in
$AdS_5\times S^5$ background", Nucl.\ Phys.\ \textbf{B533} (1998)
109, \href{http://arxiv.org/abs/hep-th/9805028}{\texttt{arXiv:hep-th/9805028}}.

\bibitem{KalloshRajaraman98}
R.~Kallosh, J.~Rahmfeld and A.~Rajaraman, "Near horizon
superspace", JHEP \textbf{9809} (1998) 002, \href{http://arxiv.org/abs/hep-th/9805217}{\texttt{arXiv:hep-th/9805217}}.

\bibitem{BBHZ}
N.~Berkovits, M.~Bershadsky, T.~Hauer, S.~Zhukov and B.~Zwiebach, "Superstring theory on $AdS_2\times S^2$ as a coset supermanifold", Nucl.\ Phys.  \textbf{B567} (2000) 61, \href{http://arxiv.org/abs/hep-th/9907200}{\texttt{arXiv:hep-th/9907200}}.

\bibitem{RoibanSiegel}
R.~Roiban and W.~Siegel, "Superstrings on $AdS_5\times S^5$ supertwistor space", JHEP \textbf{0011} (2000) 024, \href{http://arxiv.org/abs/hep-th/0010104}{\texttt{arXiv:hep-th/0010104}}.

\bibitem{AF08}
G.~Arutyunov and S.~Frolov, "Superstrings on $AdS_4\times
CP^3$ as a Coset Sigma-model", JHEP \textbf{0809} (2008) 129, \href{http://arxiv.org/abs/0806.4940}{\texttt{arXiv:0806.4940 [hep-th]}}.

\bibitem{Stefanski}
B.J.~Stefanski, "Green-Schwarz action for Type IIA
strings on $AdS_4\times CP^3$", Nucl.\ Phys.\ \textbf{B808} (2009)
80, \href{http://arxiv.org/abs/0806.4948}{\texttt{arXiv:0806.4948 [hep-th]}}.

\bibitem{PS}
P.~Fre and P.A.~Grassi, "Pure Spinor Formalism for
$OSp(N|4)$ backgrounds", \href{http://arxiv.org/abs/0807.0044}{\texttt{arXiv:0807.0044 [hep-th]}}.\\
G.~Bonelli, P.A.~Grassi and H.~Safaai, "Exploring pure spinor string theory on $AdS_4\times\mathbb{CP}^3$", JHEP \textbf{0810} (2008) 085, \href{http://arxiv.org/abs/0808.1051}{\texttt{arXiv:0808.1051 [hep-th]}}.\\
R.~D'Auria, P.~Fre, P.A.~Grassi and M.~Trigiante,
"Superstrings on $AdS_4\times CP^3$ from Supergravity", Phys. Rev. \textbf{D79} (2009) 086001,
\href{http://arxiv.org/abs/0808.1282}{\texttt{arXiv:0808.1282 [hep-th]}}.

\bibitem{GSWnew}
J.~Gomis, D.~Sorokin and L.~Wulff, "The complete $AdS_4\times
CP^3$ superspace for type IIA superstring and $D-$branes", JHEP
\textbf{0903} (2009) 015, \href{http://arxiv.org/abs/0811.1566}{\texttt{arXiv:0811.1566 [hep-th]}}.\\
P.A.~Grassi, D.~Sorokin and L.~Wulff, "Simplifying superstring and
$D-$brane actions in $AdS_4\times\mathbb{CP}^3$ superbackground", JHEP
\textbf{0908} (2009) 060,
\href{http://arxiv.org/abs/0903.5407}{\texttt{arXiv:0903.5407
[hep-th]}}.

\bibitem{DHIS}
M.J.~Duff, P.S.~Howe, T.~Inami and K.S.~Stelle, "Superstrings in
$D=10$ from supermembranes in $D=11$", Phys. Lett. \textbf{B191}
(1987) 70.\\
P.S.~Howe and E.~Sezgin, "The supermembrane revisited", Class. Quantum Grav. \textbf{22} (2005) 2167, \href{http://arxiv.org/abs/hep-th/0412245}{\texttt{arXiv:hep-th/0412245}}.

\bibitem{Plefka98}
B.~de Wit, K.~Peeters, J.~Plefka and A.~Sevrin, "The M-theory two-brane in $AdS_4\times S^7$ and $AdS_7\times S^4$", Phys. Lett. \textbf{B443} (1998) 143, \href{http://arxiv.org/abs/hep-th/9808052}{\texttt{arXiv:hep-th/9808052}}.

\bibitem{Pope}
B.E.W.~Nilsson and C.~Pope, "Hopf fibration of eleven dimensional supergravity", Class. Quantum Grav. \textbf{1} (1984) 499.

\bibitem{STV85}
D.P.~Sorokin, V.I.~Tkach and D.V.~Volkov, "Kaluza-Klein theories
and spontaneous compactification mechanisms of extra space
dimensions", \emph{In *Moscow 1984, Proceedings, Quantum Gravity*,
376-392}.\\
D.P.~Sorokin, V.I.~Tkach and D.V.~Volkov, "On the
relationship between compactified vacua of $D=11$ and $D=10$
supergravities", Phys. Lett. \textbf{B161} (1985) 301.

\bibitem{SW1009}
D.~Sorokin and L.~Wulff, "Evidence for the classical integrability of the complete $AdS_4\times\mathbb{CP}^3$ superstring", JHEP \textbf{1011} (2010) 143, \href{http://arxiv.org/abs/1009.3498}{\texttt{arXiv:1009.3498 [hep-th]}}.

\bibitem{Zarembo}
J.A.~Minahan and K.~Zarembo, "The Bethe-ansatz for $\mathcal N = 4$ super Yang-Mills", JHEP \textbf{0303} (2003) 013,  \href{http://arxiv.org/abs/hep-th/0212208}{\texttt{arXiv:hep-th/0212208}}.

\bibitem{BPR}
I.~Bena, J.~Polchinski and R.~Roiban, "Hidden symmetries of the $AdS_5\times S^5$ superstring", Phys.\ Rev. \textbf{ D69} (2004) 046002, \href{http://arxiv.org/abs/hep-th/0305116}{\texttt{arXiv:hep-th/0305116}}.

\bibitem{JPA}
J. Phys. A \textbf{42} (2009) N25: special issue on "Gauge-string duality and integrability: progress and outlook", eds. C.~Kristjansen, M.~Staudacher and A.~Tseytlin.

\bibitem{U08}
D.V.~Uvarov, "$AdS_4\times\mathbb{CP}^3$ superstring and $D=3$
$\mathcal N=6$ superconformal symmetry", Phys. Rev. \textbf{D79}
(2009) 106007, \href{http://arxiv.org/abs/0811.2813}{\texttt{arXiv:0811.2813 [hep-th]}}.

\bibitem{9807115}
G.~Dall'Agata, D.~Fabbri, C.~Fraser, P.~Fre, P.~Termonia and M.~Trigiante, "The $OSp(8|4)$ singleton action from the supermembrane", Nucl. Phys. \textbf{B542} (1999) 157, \href{http://arxiv.org/abs/hep-th/9807115}{\texttt{arXiv:hep-th/9807115}}.

\bibitem{Kallosh2}
R.~Kallosh, "Superconformal actions in Killing gauge", \href{http://arxiv.org/abs/hep-th/9807206}{\texttt{arXiv:hep-th/9807206}}.

\bibitem{PST}
P.~Pasti, D.P.~Sorokin and M.~Tonin, "On gauge-fixed superbrane actions in AdS superbackgrounds", Phys.\ Lett.\  \textbf{B447} (1999) 251, \href{http://arxiv.org/abs/hep-th/9809213}{\texttt{arXiv:hep-th/9809213}}.

\bibitem{MT2000}
R.R.~Metsaev and A.A.~Tseytlin, "Superstring action in $AdS_5\times S^5$: $\kappa-$symmetry light cone gauge", Phys.\ Rev. \textbf{D63} (2001) 046002, \href{http://arxiv.org/abs/hep-th/0007036}{\texttt{arXiv:hep-th/0007036}}.\\
R.R.~Metsaev, C.B.~Thorn and A.A.~Tseytlin, "Light-cone
superstring in AdS space-time", Nucl.\ Phys.\ \textbf{B596} (2001)
151, \href{http://arxiv.org/abs/hep-th/0009171}{\texttt{arXiv:hep-th/0009171}}.

\bibitem{Benna}
M. Benna, I. Klebanov, T. Klose and M. Smedback, "Superconformal Chern-Simons
theories and $AdS_4/CFT_3$ correspondence", JHEP \textbf{0809} (2008) 072, \href{http://arxiv.org/abs/0806.1519}{\texttt{arXiv:0806.1519 [hep-th]}}.

\bibitem{Cederwall}
M.~Cederwall, "Superfield actions for $\mathcal N=8$ and $\mathcal N=6$ conformal theories in three dimensions", JHEP \textbf{0810} (2008) 070, \href{http://arxiv.org/abs/0809.0318}{\texttt{arXiv:0809.0318 [hep-th]}}.

\bibitem{Buchbinder}
I.L.~Buchbinder, E.A.~Ivanov, O.~Lechtenfeld, N.G.~Pletnev, I.B.~Samsonov and B.M.~Zupnik, "ABJM models in $\mathcal N=3$ harmonic superspace", JHEP \textbf{0903} (2009) 096, \href{http://arxiv.org/abs/0811.4774}{\texttt{arXiv:0811.4774  [hep-th]}}.

\bibitem{Claus}
P.~Claus and R.~Kallosh, "Superisometries of the $AdS\times S$ superspace", JHEP \textbf{9903} (1999) 014, \href{http://arxiv.org/abs/hep-th/9812087}{\texttt{arXiv:hep-th/9812087}}.

\bibitem{Robins}
P.~Claus, J.~Rahmfeld, H.~Robins, J.~Tannenhauser and Y.~Zunger, "Isometries in anti-de Sitter and conformal superspaces", JHEP \textbf{0007} (2000) 047, \href{http://arxiv.org/abs/hep-th/0007099}{\texttt{arXiv:hep-th/0007099}}.

\bibitem{GSorig}
M.B.~Green and J.H.~Schwarz, "Covariant Description Of Superstrings", Phys.\ Lett.\ \textbf{B136} (1984) 367.\\
 M.B.~Green and J.H.~Schwarz, "Properties Of The Covariant Formulation Of Superstring Theories", Nucl.\ Phys.\ \textbf{B243} (1984) 285.

\bibitem{Hatsuda}
M.~Hatsuda and M.~Sakaguchi, "Wess-Zumino term for $AdS$ superstring", Phys.\ Rev.\ \textbf{D66} (2002) 045020, \href{http://arxiv.org/abs/hep-th/0205092}{\texttt{arXiv:hep-th/0205092}}.

\bibitem{VA72}
D.V.~Volkov and V.P.~Akulov, "Possible universal neutrino interaction", JETP Lett. \textbf{16} (1972) 438.\\
D.V.~Volkov and V.P.~Akulov, "Is the Neutrino a Goldstone Particle?", Phys.\ Lett. \textbf{B46} (1973) 109.

\bibitem{BLS}
M.~Bandres, A.~Lipstein and J.H.~Schwarz, "Studies of the ABJM Theory in a Formulation with Manifest $SU(4)$ $R-$Symmetry", JHEP \textbf{0809} (2008) 027, \href{http://arxiv.org/abs/0807.0880}{\texttt{arXiv:0807.0880  [hep-th]}}.

\bibitem{semiclass}
Bin Chen and Jun-Bao Wu, "Semi-classical strings in $AdS_4\times\mathbb{CP}^3$", JHEP \textbf{0809} (2008) 096,
\href{http://arxiv.org/abs/0807.0802}{\texttt{arXiv:0807.0802 [hep-th]}}.\\
T.~McLoughlin and R.~Roiban, "Spinning strings at one-loop in $AdS_4\times\mathbb{P}^3$", JHEP \textbf{0812} (2008) 101, \href{http://arxiv.org/abs/0807.3965}{\texttt{arXiv:0807.3965 [hep-th]}}.\\
L.F.~Alday, G.~Arutyunov and D.~Bykov, "Semiclassical quantization of spinning strings in $AdS_4\times\mathbb{CP}^3$", JHEP \textbf{0811} (2008) 089, \href{http://arxiv.org/abs/0807.4400}{\texttt{arXiv:0807.4400 [hep-th]}}.\\
C.~Krishnan, "AdS4/CFT3 at one loop", JHEP \textbf{0809} (2008) 092, \href{http://arxiv.org/abs/0807.4561}{\texttt{arXiv:0807.4561 [hep-th]}}.\\
T.~McLoughlin, R.~Roiban and A.A.~Tseytlin, "Quantum spinning strings in $AdS_4\times\mathbb{CP}^3$: testing the Bethe ansatz proposal", JHEP \textbf{0811} (2008) 069, \href{http://arxiv.org/abs/0809.4038}{\texttt{arXiv:0809.4038 [hep-th]}}.

\bibitem{Bykov}
D.~Bykov, "Symmetry Algebra Of The $AdS_4\times\mathbb{CP}^3$ Superstring", Theor.\ Math.\ Phys. \textbf{163} (2010) 496, \href{http://arxiv.org/abs/0904.0208}{\texttt{arXiv:0904.0208 [hep-th]}}.

\end{thebibliography}
\end{document}